\documentclass[12pt, oneside]{article}   	% use "amsart" instead of "article" for AMSLaTeX format
\usepackage[top=1in, bottom=1in, left=1in, right=1in]{geometry}
\usepackage[parfill]{parskip}    		% Activate to begin paragraphs with an empty line rather than an indent
\usepackage{graphicx}				% Use pdf, png, jpg, or eps§ with pdflatex; use eps in DVI mode
\usepackage{amssymb}
\usepackage{amsmath,mathtools}				% for aligning and spliting equations
\usepackage{hyperref}				% for link to url
\usepackage{placeins}			%for float barriers
\usepackage{authblk}

\makeatletter
\renewcommand\AB@affilsepx{ -- \protect\Affilfont}
\makeatother

\usepackage[margin=1cm]{caption} % For caption size

\usepackage{titlesec} % To make the section spacing and title more compact
\titlespacing*{\section}{0pt}{0.1ex plus .1ex minus .1ex}{.1ex minus .1ex}
\titlespacing*{\subsection}{0pt}{0.1ex plus .1ex minus .1ex}{.1ex minus .1ex}
\titleformat{\section}{\bfseries}{\thesection}{1em}{}
\titleformat{\subsection}{\bfseries}{\thesubsection}{1em}{}

\usepackage{setspace}

\usepackage{natbib}
\usepackage{lineno}
\title{\large State-space models' dirty little secrets: even simple linear Gaussian models can have estimation problems}
\author[1,*]{\normalsize Marie Auger-M\'eth\'e}
\author[1]{Chris Field}
\author[2]{Christoffer M. Albertsen}
\author[3]{Andrew E. Derocher}
\author[3,4]{Mark A. Lewis}
\author[5]{Ian D. Jonsen}
\author[1]{Joanna Mills Flemming}
\affil[1]{\small Dalhousie University, Department of Mathematics and Statistics, Halifax, B3H 4R2, Canada}
\affil[2]{Technical University of Denmark, National Institute of Aquatic Resources, Charlottenlund, 2920, Denmark}
\affil[3]{University of Alberta, Department of Biological Sciences, Edmonton, T6G 2E9, Canada}
\affil[4]{University of Alberta, Department of Mathematical and Statistical Sciences, Edmonton, T6G 2G1, Canada}
\affil[5]{Macquarie University, Department of Biological Sciences, Sydney, 2109, Australia}

\affil[*]{auger-methe@dal.ca}
\date{}

\begin{document}
\begin{spacing}{1.9}

\maketitle

\begin{abstract}
State-space models (SSMs) are increasingly used in ecology to model time-series such as animal movement paths and population dynamics. This type of hierarchical model is often structured to account for two levels of variability: biological stochasticity and measurement error. SSMs are flexible. They can model linear and nonlinear processes using a variety of statistical distributions. Recent ecological SSMs are often complex, with a large number of parameters to estimate. Through a simulation study, we show that even simple linear Gaussian SSMs can suffer from parameter- and state-estimation problems. We demonstrate that these problems occur primarily when measurement error is larger than biological stochasticity, the condition that often drives ecologists to use SSMs. Using an animal movement example, we show how these estimation problems can affect ecological inference. Biased parameter estimates of a SSM describing the movement of polar bears (\textit{Ursus maritimus}) result in overestimating their energy expenditure. We suggest potential solutions, but show that it often remains difficult to estimate parameters. While SSMs are powerful tools, they can give misleading results and we urge ecologists to assess whether the parameters can be estimated accurately before drawing ecological conclusions from their results. 
\end{abstract}

\textbf{Keywords:} Ocean Tracking Network, State-space model, Parameter estimability, Dynamic linear model

\section{Introduction}

State-space models (SSMs) are increasingly used in ecology and are becoming the favoured statistical framework for modelling animal movement and population dynamics \citep{Buckland2004,Patterson2008,McClintock2012,Newman2014}. SSMs are desirable because they are structured so as to differentiate between two distinct sources of variability: the biological or process variation (e.g., demographic stochasticity) and the measurement error associated with the sampling method \citep{Patterson2008,Newman2014}. Because marine observations are often associated with large measurement errors that can mask biological signals, much of the early development of SSMs in ecology was by marine ecologists and fisheries scientists \citep[e.g.,][]{Newman1998,Jonsen2003,Sibert2003}. The SSM framework has since become a general approach to account for multiple levels of stochasticity when modelling time-series, making them increasingly popular in the terrestrial literature \citep[e.g.,][]{Csillery2013,Fukasawa2013,Flesch2014}. Here, we demonstrate that even simple SSMs can be problematic. The model we chose is often used to explain how SSMs can account for two levels of stochasticity \citep[e.g.,][]{Newman2014}, yet, we show that it suffers from parameter- and state-estimation problems.

SSMs are a type of hierarchical model, in which one level treats the underlying unobserved states as an autocorrelated process, while another level accounts for measurement error \citep{Cressie2009}. The SSM framework is flexible, especially when fitted with Monte Carlo methods such as particle filters or Markov Chain Monte Carlo (MCMC). SSMs can be used to model a variety of linear and nonlinear processes, and can represent stochasticity with diverse statistical distributions \citep[e.g.,][]{Pedersen2011,McClintock2012, Albertsen2015}. The flexibility of the SSM approach allows ecologists to build complex models that describe the biological and measurement processes with levels of detail that were previously unattainable. 

While the SSM framework is flexible, much of its theoretical foundation is based on simple linear Gaussian SSMs \citep[sometimes referred as normal dynamic linear models, see][]{Newman2014}. An example of a simple univariate linear Gaussian SSM is the one we will use to demonstrate parameter-estimability problems: 
\begin{linenomath}
\begin{align}
	\text{Measurement eq} && y_t &= x_t + \epsilon_t \,, & \epsilon_t \sim N(0, \sigma_{\epsilon}^2) \, & \text{, where } \, t \geq 1, \sigma_{\epsilon}^2 >0 \label{eq:mesEq}\\
	\text{Process eq} && x_t &=  \rho x_{t-1} + \eta_t \,, & \eta_t \sim N(0, \sigma_{\eta}^2) \, & \text{, where } \, t \geq 1, \sigma_{\eta}^2 >0 \,, -1 < \rho < 1, \label{eq:procEq}
\end{align}
\end{linenomath}
where $\mathbf{y} = (y_1, y_2, ..., y_t, ..., y_n) $ are observed at regular time intervals $\mathbf{t} = (1, ..., n)$ for a time-series of length $n$ and $\mathbf{x} = (x_0, x_1, ..., x_t, ..., x_n)$ are the true unobserved states, with $x_0$ representing the initial state. An ecological example of such a time-series would be a series of yearly population size estimates. For instance, \citet{Newman2014} use this model to introduce SSM for population dynamics with $x_t$ representing the true but unknown abundance of an animal population at time $t$, $y_t$ an unbiased observation of the population size at time $t$, and $\rho$ the population growth rate. 

The origin of SSMs is intimately linked with the Kalman filter, a recursive procedure to estimate the unobserved states based on inaccurate observations (e.g., estimating the true fish abundance based on catch data). The Kalman filter was developed to estimate states based on a model without unknown parameter values \citep{Kalman1960}. However, in ecological applications, most parameters need to be estimated \citep[e.g.,][]{McClintock2012}. Fitting methods for SSMs, such as the Kalman filter, are now used to facilitate both state and parameter estimation \citep{Johnson2008}. In many cases, SSMs are used to estimate variance parameters because they are designed to differentiate measurement error from process stochasticity \citep{Dennis2006,Simmons2015}. While estimating parameters is often a means to estimate the unobserved states \citep[e.g.,][]{Johnson2008,Albertsen2015}, parameters themselves can be of interest because they describe the underlying dynamics of the system, or behaviour of the animal \citep[e.g.][]{MillsFlemming2010, McClintock2012}.

Estimability problems associated with SSMs and other hierarchical models have been discussed in the population dynamics literature \citep[e.g.,][]{Dennis2006,Knape2008}. In particular, previous studies have emphasized how difficult it is to use SSMs to estimate density dependence parameters \citep{Knape2008,Polansky2009} and to differentiate process stochasticity from measurement error \citep[e.g.,][]{Dennis2006}. However, the existence of parameter estimation problems have been largely overlooked in the movement literature, and by those that use complex Bayesian SSMs. As SMMs are becoming the favoured framework for many ecological analyses \citep{Buckland2004,Patterson2008,McClintock2012,Newman2014}, and are gaining popularity in other fields \citep[e.g.,][]{Cao2014}, it is timely to warn researchers of their weaknesses.

Here, we use simulations to show that simple SSMs can have severe parameter-estimability problems that in turn affect state estimates. These problems are more frequent when the measurement error is large, the very condition under which SSMs are needed, and can persist even when we incorporate measurement error information. While our main estimation approach consists of maximizing the likelihood numerically through Template Model Builder (TMB; developed by Kasper Kristensen and available at www.tmb-project.org), we show that these problems persist across a wide range of platforms and statistical frameworks, including when the parameters and states are estimated via Bayesian methods. We use the polar bear (\textit{Ursus maritimus}) movement data that led us to notice these problems to demonstrate the effect of estimation problems on the biological interpretation of results. Finally, we discuss techniques to diagnose and, when possible, alleviate estimability problems.

\section{Methods}

\subsection{Demonstration of the problem}
\label{sec:sim}

When we fit models to data, we want the parameters to be identifiable, which means that, given perfect data (e.g., an infinitely long time-series), it is possible to learn the true values of parameters. Assessing parameter identifiability is often difficult and a more attainable goal is to assess estimability. Estimability means that, given the data at hand, the method used to approximate the parameter yields a unique estimate. When the maximum value of the likelihood function occurs at more than one parameter value, the parameter is nonestimable. The quality of parameter estimates can be assessed in terms of: its variance, measured over multiple repeated estimations; bias, the expected difference between the estimate and true value of the parameter; or mean square error, a composite of bias and variance. To demonstrate that the estimates of the parameters and states of SSMs can be inaccurate, we simulated a set of time-series using the model presented in eq. \ref{eq:mesEq}-\ref{eq:procEq}. In all simulations,  the values for the initial state, $x_0$, the measurement error, $\sigma_{\epsilon}$, and the correlation, $\rho$, were set to 0, 0.1, and 0.7, respectively. In Appendix \ref{App:Rho1} (Supplementary information), we explored other $\rho$ values, including a simpler model where $\rho$ is fixed to 1. Note that while this simpler model has fewer parameters to estimate, it is no longer stationary \citep{Durbin2001}. To investigate whether the ratio of measurement to process stochasticity affected estimation, we simulated a range of $\sigma_{\eta}$ values: $(0.01,0.02,0.05,0.1,0.2,0.5,1)$. For each parameter set, we simulated 200 time-series each with 100 observations ($n=100$). Analyses using longer time-series (n=500) are presented in Appendix \ref{App:n500} (Supplementary information).

For each simulation, we estimated the parameters, $\boldsymbol{\theta}=(\sigma_{\epsilon}, \rho, \sigma_{\eta})$, and states, $\mathbf{x}$, using the \texttt{R} \citep{R2015} package \texttt{TMB}. This \texttt{R} package is similar to AD Model Builder \citep{Fournier2012} in that it uses automatic differentiation and the Laplace approximation. Finding the Maximum Likelihood Estimate (MLE) of the parameters of a SSM requires the maximization of the marginal distribution of the observations \citep{Newman2014}. For the model presented in eq. \ref{eq:mesEq}-\ref{eq:procEq}, this involves maximizing  the following likelihood:
\begin{linenomath}
\begin{align}
L_{\theta}(\sigma_{\epsilon},\rho,\sigma_{\eta}|\mathbf{y})	&= \int \prod_{t=1}^{n}{	p(y_t | x_t) p(x_t|x_{t-1})} d\mathbf{x},  \label{eq:likGen}\\
&= \int \prod_{t=1}^{n}{N(x_t, \sigma_{\epsilon}^2) N(\rho x_{t-1}, \sigma_{\eta}^2)} d\mathbf{x} \label{eq:lik}.
\end{align}
\end{linenomath}
To get the marginal distribution, we integrate over the states, $\mathbf{x}=(x_1,...,x_n)'$. In \texttt{TMB}, this integration is achieved using the Laplace approximation, which in turn also returns state estimates \citep{Albertsen2015}. While we refer to state ``estimation'',  this process is sometimes called ``prediction'' because states can be interpreted as random variables \citep{Newman2014}. In this example, we assumed that the initial state is known (i.e., $x_0 = 0$), which should help the estimation process. In instances where the initial state value is unavailable, the initial state can be modelled as $x_0 \sim N(\mu, \sigma_0)$ \citep{Durbin2001}. 
\texttt{TMB} calculates standard errors for the estimated parameters by using the inverse of the observed Fisher information, i.e. the Hessian of the log likelihood \citep[similar to ADMB, see ][]{Fournier2012}. To calculate the $95\%$ confidence intervals (CI), we multiplied the aforementioned standard errors by the 2.5 and 97.5\textsuperscript{th} percentiles of the normal distribution \citep[i.e., the quadratic approximation in][]{Bolker2008}. 

To demonstrate that the problem is widespread across different statistical platforms, we also fitted the simulated data using two popular \texttt{R} packages: \texttt{dlm} \citep{Petris2010} and \texttt{rjags} \citep{Plummer2014}. \texttt{dlm} uses the Kalman filter for the state estimates and calculates the MLE with numerical optimization methods. \texttt{rjags} is an \texttt{R} interface to JAGS, a program that can be used to fit Bayesian hierarchical models using MCMC methods (Supplementary information: Appendix \ref{App:DLMandJAGS}).

We evaluated the parameter-estimation performance of SSMs by comparing the estimated and simulated values. Similar to \citet{Pedersen2011}, we evaluated the state-estimation performance with the root mean square error (RMSE):
\begin{linenomath}
\begin{align}
	\text{RMSE} &= \sqrt{\frac{1}{n} \sum_{t=1}^{n}{(\hat{x}_{t} - x_{t})^2}},
\end{align}
\end{linenomath}
where $\hat{x}_{t}$ is the estimated state at time $t$ and $x_{t}$ is the simulated (i.e., true) state at time $t$. To assess whether the state-estimation performance was affected by the parameter-estimation problems, we compare $\text{RMSE}_{\hat{\theta}}$, for which the parameters, $\boldsymbol{\theta} = (\sigma_{\epsilon},\rho, \sigma_{\eta})$, were also estimated, to $\text{RMSE}_{\theta}$, for which the parameter values were fixed at the values used to simulate the data.

To investigate the potential causes of the parameter-estimation problem, we explored the likelihood profile for a subset of the problematic simulations. We used the same simulations and parameter values as above, with the exception that we only examined the most problematic values: $\sigma_{\eta}=(0.01,0.02,0.05)$ (see Results). Because they are associated with high measurement error to process stochasticity ratios, these values also represent the conditions when SSMs are most needed. For each scenario (i.e., different values of $\sigma_{\eta}$), we randomly chose one simulation for which the $\text{RMSE}_{\hat{\theta}}$ was $50\%$ larger than $\text{RMSE}_{\theta}$. Again, we used \texttt{TMB} to estimate parameter values, $\boldsymbol{\theta}$, and the states, $\mathbf{x}$. To examine whether the estimation problems were associated with the simultaneous estimation of states and parameters, we estimated parameters when the state values were fixed to their simulated values (Supplementary information: Appendix \ref{App:FixX}). As a final investigation of the causes of the estimation problems, we show how these problems are associated with known limitations of the autoregressive-moving-average (ARMA) models (Supplementary information: Appendix \ref{App:ARMA}).

\subsection{Incorporating measurement error information}
\label{MesMethod}

Many ecologists incorporate information on measurement error in their model by either fixing parameter values or, in a Bayesian framework, using informative priors \citep[e.g.][]{Jonsen2003,Jonsen2005,Johnson2008}. We investigated whether fixing the measurement error resolved the parameter estimation problem. To do so, we fitted our simple likelihood (eqn. \ref{eq:lik}) to the same simulations, but we fixed the standard deviation of the measurement equation to the value used to simulate the data, $\sigma_{\epsilon} = 0.1$. We only estimated the remaining parameters, $\boldsymbol{\theta}_{m} = (\rho, \sigma_{\eta})$. As above, we investigated the parameter estimates, RMSE of the states, and likelihood profiles.

\subsection{Ecological example}

The movement of many animals, such as birds, fish and marine mammals, is a combination of the voluntary movement of the animal (active movement) and drift (passive displacement resulting from ocean or wind currents). Currents do not always direct animals towards their goals, and moving against currents may require a substantial amount of energy \citep[e.g.,][]{Weimerskirch2000}. To understand how currents affect the behavioural strategies of an animal, it is necessary to distinguish between the voluntary movement of the animal and drift \citep{Gaspar2006}. The voluntary movement can then be used as a proxy of energy expenditure, or can be integrated into an energy budget model to assess the effects of movement on survival and reproduction \citep{Gaspar2006,Molnar2010, Molnar2014}. While developments in satellite telemetry are providing increasingly precise measurements of animal movement paths, it is difficult to differentiate between drift and voluntary movement because wind, ocean, and sea ice drift data are often associated with large errors \citep[e.g.,][]{Schwegmann2011,Fossette2012}. 

We noticed the estimation problems of linear Gaussian SSMs when developing a model that would differentiate between the voluntary movement of polar bears and sea ice drift. Polar bears often move in the reverse direction of the sea ice drift \citep{Mauritzen2003,Auger-Methe2015} and sea ice drift can be associated with large errors \citep{Schwegmann2011}. As a proxy of energy expended by bears, we wanted to estimate the voluntary movement. As a first test, we developed a 2 dimensional SSM  that accounts for error in ice drift data:
\begin{linenomath}
\begin{align}
	\text{Initial state} && \mathbf{x}_0 &\sim N(\boldsymbol{\mu},\mathbf{P}_0) & \label{eq:pbX0} \\
	\text{Measurement eq} && \mathbf{y}_t &= \mathbf{x}_t + \mathbf{s}_t + \boldsymbol{\epsilon}_t, &\boldsymbol{\epsilon}_t \sim N(0,\mathbf{H}) \label{eq:pbObs}\\
	\text{Process eq} && \mathbf{x}_t &= \mathbf{T} \mathbf{x}_{t-1} + \boldsymbol{\eta}_t, &\boldsymbol{\eta}_t \sim N(0,\mathbf{Q}), \label{eq:pbPro}
\end{align}
\end{linenomath}
where $\mathbf{y}_t = \left[ \begin{smallmatrix} y_{t,u} \\ y_{t,v} \end{smallmatrix} \right]$ is the measured daily displacement of the polar bear based on the GPS collar data, $\mathbf{x}_t = \left[ \begin{smallmatrix} x_{t,u} \\ x_{t,v} \end{smallmatrix} \right]$  is the voluntary displacement of the polar bear, and $\mathbf{s}_t = \left[ \begin{smallmatrix} s_{t,u} \\ s_{t,v} \end{smallmatrix} \right]$ is the daily sea ice drift experienced by the bear. Here, the measurement error, $ \boldsymbol{\epsilon}_t$, is associated with the ice data, not the polar bear location data. The location data were determined by GPS, for which the error is negligible \citep[$<30m$, see][]{Tomkiewicz2010}. For simplicity, we assumed that the two geographic coordinates are independent, thus:
\begin{linenomath}
\begin{align}
	\mathbf{P_0} &=  
		\begin{bmatrix} 
			\sigma^2_{0} & 0\\
			0 & \sigma^2_{0} 
		\end{bmatrix}, &
	\mathbf{H} &=  
		\begin{bmatrix} 
			\sigma^2_{\epsilon,u} & 0\\
			0 & \sigma^2_{\epsilon,v} 
		\end{bmatrix}, &
	\mathbf{Q} &= 
		\begin{bmatrix} 
			\sigma^2_{\eta,u} & 0\\
			0 & \sigma^2_{\eta,v} 
		\end{bmatrix}, &
	\mathbf{T} &= 
		\begin{bmatrix} 
			\rho_{u} & 0\\
			0 & \rho_{v} 
		\end{bmatrix}.
\end{align}
\end{linenomath}
Because eq. \ref{eq:pbX0}-\ref{eq:pbPro} model displacements, the elements of $\mathbf{H}$ represent the measurement error in the sea ice drift data and those of $\mathbf{Q}$ are associated with the speed of the bear. Similar to $\gamma$ in \citet{Jonsen2005}, $\rho_u$ and $\rho_v$ represent the degree of autocorrelation in the random walk. To initialize the model we used $\mathbf{\mu} = \left[ \begin{smallmatrix} 0 \\0 \end{smallmatrix} \right]$ and $\sigma^2_{0}=15$. We chose 15 km as it is the standard deviation of the observed daily displacements of the polar bears in the $u$- and $v$-direction. 

We used the daily movement of 15 polar bears collared in the Beaufort Sea in the spring of 2007-2011. The bears were immobilized with standard methods \citep{Stirling1989} and equipped with Telonics Inc. (Mesa, AZ) collars. All capture and  handling procedures were carried out in accordance with the protocols approved by the University of Alberta Animal Care and Use Committee for Biosciences. We used the Polar Pathfinder Daily 25km Ease-Grid Sea Ice Motion Vectors \citep{Fowler2003}, which are daily estimates of sea ice displacements in the $u$- and $v$-directions of the Northern Hemisphere azimuthal equal-area EASE-Grid projection developed for polar sea ice data  \citep{Brodzik2002}. We used the same movement data and data handling procedures as in \citet{Auger-Methe2015}, including interpolating the ice drift data at each bear location, assigning a drift value of zero for landfast ice, and excluding the three days after collaring to remove movements affected by handling. The only differences in the data used here, are that we excluded all bears that spent time on land and considered days with missing sea ice data as missing observations (i.e., we considered both $\mathbf{y}_t$ and $\mathbf{s}_t$ as missing that day).

Our goal was to use the SSM to estimate the energy expenditure of each bear. Our proxy was the total voluntary bear displacement: 
\begin{linenomath}
\begin{align}
	d = \sum^{n}_{t=1}{\sqrt{\hat{x}^2_{t,u} + \hat{x}^2_{t,v}}} \label{eq:disp},
\end{align}
\end{linenomath}
where $\hat{x}_{t,u}$ and $\hat{x}_{t,v}$ are the estimates of the daily voluntary bear displacements in the $u$- and $v$-directions. The number of days, $n$, included in the time-series will affect our estimate of $d$. For consistency, we set $n$ to be 342, the length of the shortest time-series across the 15 bears. To assess the effects of estimation problems on our ecological interpretation, we simulated movement paths similar to those described by the polar bear data (Supplementary information: Appendix \ref{App:PBsim}).

The code is available at \url{https://gitlab.oceantrack.org/otn-statistical-modelling-group/SSMestProblems} and as Supplementary data.

\section{Results}
\subsection{Simulations results}
\label{SimRes}

According to the simulation results, parameter estimation was often inaccurate, and these problems affected the state estimates (Fig. \ref{fig:TMBParEstSum}). The parameter estimates were often far from their true values, and their distributions often bimodal (Fig. \ref{fig:TMBParEstSum}, Supplementary Fig. \ref{fig:TMBParEst}). In many cases, the estimates for $\sigma_{\epsilon}$ and $\rho$ had peaks close to 0. The $\text{RMSE}_{\hat{\theta}}$ of the state estimates had either a bimodal distribution, or a long tail compared to that of the $\text{RMSE}_{\theta}$ (Supplementary Fig. \ref{fig:TMBParEst}). In other words, when the parameters were estimated, many replicates had much higher state estimate error than when the true parameter values were used (Fig. \ref{fig:TMBParEstSum}). In fact, 29.6\% of the simulations had a $\text{RMSE}_{\hat{\theta}}$ value that was $50\%$ larger than their $\text{RMSE}_{\theta}$. When the simulations had high measurement error to process stochasticity ratios, the estimation problems for the states and two biologically relevant parameters, ($\rho, \sigma_{\eta}$) were much higher (Fig. \ref{fig:TMBParEstSum}). The $\text{RMSE}_{\hat{\theta}}$ in some of these cases was close to 10 times greater than the simulated process stochasticity.

Our supplementary analyses demonstrated that similar estimation problems occurred when \texttt{dlm} and \texttt{rjags} were used (Supplementary information: Appendix \ref{App:DLMandJAGS}). However, while the parameters estimated with \texttt{rjags} were often biased, their distributions did not contain a peak at 0. Increasing the length of the time-series improved parameter and state estimation (Supplementary information: Appendix \ref{App:n500}). However, 500 time steps were insufficient to completely eliminate problems. Our supplementary analyses also show that the problems are less apparent when $\rho$ is close to 1, or when we used the simpler non-stationary local-level model, which fixes the value of $\rho=1$ (Supplementary information: Appendix \ref{App:Rho1}).

The likelihood profiles of a subset of the problematic simulations revealed that the likelihood was flat in some areas and sometimes bimodal or jagged (Fig. \ref{fig:LProfAll}). The CI of many parameters excluded the true simulated value. Because the estimated measurement error of these simulations were close to 0, the estimated states were very close to the observations and far from their true simulated values (Fig. \ref{fig:LProfAll}D,H,L). When the states were fixed to their simulated rather than estimated values, the likelihood profiles were unimodal and most CI included the true parameter values, indicating that the problem lies in simultaneously estimating the states and the parameters (Supplementary information: Appendix \ref{App:FixX}).

\subsection{Fixing the measurement error}
\label{MesRes}

Fixing the standard deviation of the measurement error to the simulated value, $\sigma_{\epsilon} = 0.1$, helped reduce the estimation problems (Supplementary information: Appendix \ref{App:mesFix}). $\text{RMSE}_{\hat{\theta}}$ values were much closer to $\text{RMSE}_{\theta}$ when the measurement error was fixed rather than estimated. In this case, only 5.0\% of the simulations had a $\text{RMSE}_{\hat{\theta}}$ value that was $50\%$ larger than their $\text{RMSE}_{\theta}$. However, fixing the measurement error did not completely resolve the estimation problems. Some parameter estimates continued to be on the boundary of parameter space and far from their simulated values. In addition, some likelihood profiles remained flat and some CIs spanned the entire parameter space (see Supplementary information: Appendix \ref{App:mesFix} for more detail).

\subsection{Ecological example}

The 15 polar bears studied used overlapping areas in the Beaufort Sea (Fig. \ref{fig:Bear}A), but their parameters estimates varied widely (Fig. \ref{fig:Bear}C-H). In particular, three individuals had much lower estimated sea ice measurement error, with either $\widehat{\sigma}_{\epsilon,u} < 0.01$ and $\widehat{\sigma}_{\epsilon,v}  < 0.01$. These three individuals had total voluntary displacement estimates that were on the higher end of the range (Fig. \ref{fig:Bear} B). These results are similar to those found when we simulated movement data similar to the real polar bear data (Supplementary information: Appendix \ref{App:PBsim}). The simulations also showed that a few individuals would have $\widehat{\sigma}_{\epsilon,u} < 0.01$ and $\widehat{\sigma}_{\epsilon,v}  < 0.01$ and that these individuals would be associated with higher values of total voluntary displacement.

\section{Discussion}

Linear Gaussian SSMs, and approximations of them, are commonly used in the ecological literature to model animal movement \citep{Jonsen2003, Johnson2008, Patterson2008} and population abundance \citep[e.g.,][]{Wilson2011,Flesch2014}. These SSMs are often used to differentiate measurement error from process stochasticity and estimate the associated variance parameters \citep[e.g.,][]{Sibert2006,Wilson2011,Flesch2014,Albertsen2015}. Our results demonstrated that simple linear Gaussian SSMs can have severe parameter- and state-estimation problems, and that these problems can affect biological inferences. According to our simulations, estimation problems were more frequent when the measurement error was much larger than the process stochasticity. In such cases, the three estimated parameters were often far from their simulated values, which in turn resulted in inaccurate state estimates. The ARMA notation shows that when the measurement error is much greater than the process stochasticity there is parameter redundancy, explaining why it is difficult to accurately estimate the parameters (Supplementary information: Appendix \ref{App:ARMA}). Our simulations showed that fixing the measurement error to its true value helped, but did not completely solve the estimation problems, especially when the fixed measurement error was relatively large. This is particularly worrisome because SSMs are most needed when the measurement error is large compared to the process stochasticity, and this is the condition under which the largest estimation problems occur. 

The estimation problems are less critical when the measurement error is much smaller than the process stochasticity. While the measurement error estimates were often close to 0, the estimates for the other parameters, and those for the states, were generally accurate.  As shown by the ARMA notation, when the measurement error is much smaller than the process stochasticity the model behaves as an AR(1) process, explaining why the measurement error estimates were often close to zero (Supplementary information: Appendix \ref{App:ARMA}). In effect, the measurement error is ignored. However, when the measurement error is negligible compared to the process stochasticity, ignoring the effect of the measurement error is less likely to affect our interpretation of the biological process.

Others have discussed estimation problems associated with fitting simple linear Gaussian SSMs. A few recent ecological studies have reported difficulties when estimating variance parameters, including variance estimates close to 0 \citep{Tittensor2014a,Simmons2015}. \citet{Dennis2006}, who transformed the stochastic Gompertz population model into a linear Gaussian SSM, noted that while the process stochasticity and measurement error parameters can be estimated, multimodal likelihood functions occur and can lead to erroneous estimates. They showed that the likelihood functions tended to have multiple peaks, including two peaks associated with either no process stochasticity or no measurement error. While these two peaks can be local maxima, \citet{Dennis2006} noted that when there is substantial measurement error, one of these modes was often the global maximum. \citet{Knape2008} extended the study of the Gompertz SSM to focus on the estimability of the density dependence parameter, an autocorrelation parameter similar to $\rho$. He found that the density dependence was generally not identifiable in the presence of unknown process variability and measurement error, especially when the strength of the density dependence was close to  0. When the measurement error was known, the strength of density dependence was estimable but the estimates often remained biased. 

By extending the range of measurement error to process stochasticity ratios beyond those explored by \citet{Dennis2006} and \citet{Knape2008}, we demonstrate that relatively high measurement error can have dramatic effects on process parameter and state estimates, even when the measurement error is known. The results of \citet{Knape2008} suggested that $\rho$ values close to 0 would result in estimability problems \citep[see also][]{Forester2007}, which is not surprising. As the process becomes less autocorrelated it is harder to differentiate it from the temporally independent measurement error, suggesting that differentiating between measurement error and process stochasticity would require a large sample size when $\rho$ is far from 1. However, our results demonstrated that estimation problems remained with relatively high autocorrelation, $\rho = (0.7, 0.99)$ and $\rho$ fixed to 1, and relatively long time-series, $n=(100,500)$ (see Supplementary information: Appendices \ref{App:Rho1}-\ref{App:n500}). These results emphasize that the parameters and states are only estimable for a narrow range of conditions. Both the analysis of the ARMA formulation of our SSM and our ecological example show that parameter estimability within linear Gaussian SSMs is a general issue, not one restricted to the stochastic Gompertz population model. In fact, these problems extend to some nonlinear SSMs. For example, some of the estimated parameters of the nonlinear population SSMs of \citet{deValpine2002} had considerable bias when measurement error was large relative to process variability, \citet{Valpine2005} showed that their advance Monte Carlo kernel likelihood method could not differentiate between the process stochasticity and measurement error of the nonlinear Schaefer population model, and \citet{Polansky2009} found similar problems in the theta-Ricker model.

Left undiagnosed, biased parameter estimates will mislead conclusions based on the problematic model parameters and may affect our interpretation of the other model parameters, the state estimates, and other derived values \citep{Cressie2009,Lele2010a}. For example, stochastic population SSMs with negatively biased estimates of the process stochasticity will underestimate extinction risk \citep{Lindley2003}. In our polar bear example, erroneous estimates of measurement error and process stochasticity biased the state estimates and proxy for energy expenditure. Thus, even if the parameter values \textit{per se} are not of interest, estimation problems need to be diagnosed because their effect on state estimates are likely to affect results of ecological importance.

The first step to avoid these biased inferences is to detect the potential for parameter estimability problems, which can be done through a variety of practical means. Our simulations demonstrated that estimates at the boundary of parameter space can be indicative of a problem. For our polar bear example, we detected the estimation problem because we had no reason to believe that the three bears with sea ice measurement error close to 0 used different sea ice than the other bears. These three bears were exposed to similar levels of sea ice drift as other bears and were not geographically or temporally isolated from them. Investigating the likelihood profile can also help detect estimation problems \citep{Dennis2006,Dennis2010,Ives2010}. Indeed, the likelihood profiles of our problematic simulations had flat sections and multiple modes. However, in a Bayesian framework, the estimation problems can be obscured by the use of vague priors, as these can smooth the likelihood and affect inference \citep{Lindley2003,Dennis2006,Lele2009,Lele2010a}. When we used JAGS to estimates parameters, we had no estimates at the boundary and the posterior distributions of most parameters were unimodal, and yet, the estimates were biased (Supplementary information: Appendix \ref{App:DLMandJAGS}). A useful way to evaluate the model's capacity to separate process and measurement error parameters, is to assess the extent of correlation between these estimates (see Supplementary information: Appendix \ref{App:diagTools} for details). In the maximum likelihood context, a plot of the likelihood surface can reveal a correlation pattern symptomatic of an identifiability issue \citep{Valpine2005,Polansky2009}. In a Bayesian context, a plot of the joint posterior samples of these two parameters can reveal similar correlation patterns (Supplementary information: Appendix \ref{App:diagTools}). While few methods have been developed to formally assess parameter identifiability problems, data cloning \citep{Lele2010,Campbell2014} and the symbolic method \citep{Cole2012,Newman2014} are promising avenues.

How can we avoid these estimability problems? In many cases, a larger sample size can help (see Supplementary information: Appendix \ref{App:n500}). In particular, \citet{Dennis2010} demonstrated that sampling replicates can substantially improve the capacity of SSMs to differentiate process stochasticity from measurement error, and that it may be advantageous to design monitoring programs with multiple replicate counts per survey rather than increasing the length of the time series (i.e., number of times the survey is conducted). However, for many observational studies, ecologists are limited in their ability to gather more data and, for movement data, it is often impossible to have replicates of location estimates. An alternative is to incorporate information on the measurement error. As we demonstrated in our simulation study, when we fix the measurement error to its true value, the estimates of the other parameters improved. While some parameter-estimation problems persisted, their effect on the state estimates diminished substantially. Similarly, \citet{Valpine2005} demonstrated that knowing the ratio of process to measurement variance would improve parameter estimates. In a Bayesian framework, specifying informative priors for the measurement error could help make the other parameters identifiable and improve the state estimates \citep[][but see \citealt{Lele2009}]{Lindley2003,Cressie2009}. Another alternative is to estimate the measurement error and process stochasticity outside of the SSM framework using the principle that the measurement error is uncorrelated over time whereas the process stochasticity is temporally correlated \citep{Dowd2011}. Estimating the measurement and process standard deviations offline reduces the number of parameters to estimate within the SSM framework. Using restricted maximum-likelihood, which treats fixed-effects parameters (e.g., $\rho$) and variance components (e.g., $\sigma^2_{\eta}$, $\sigma^2_{\epsilon}$) differently, can also be valuable to remove bias in SSM estimates \citep{Dennis2010}. When the estimation problem results in variance estimate close to 0, one can limit the estimate to interior (non-zero) solutions \citep{Dennis2006,Knape2008}. In particular, \citet{Dennis2006} suggested trying a variety of starting values for the optimizer used to numerically maximize the likelihood and eliminating all solutions that involve variance with near 0 values, even if one of these is the global maximum. Finally, restructuring the model can help reduce the problem. For example, in the polar bear example, we could create a population model with a single measurement error parameter for all bears. Even if the process variability continues to differ between individuals, using one measurement error term for all bears significantly decreases the number of parameters to estimate and increases the amount of data with which the measurement error term is estimated. As a general rule decreasing the number of parameters to estimate and increasing the amount of data will help reduce estimability problems.

Not all parameters are equally affected by estimation problems. \citet{Forester2007}, who developed a linear Gaussian SSM for animal movement, demonstrated that coefficient parameters associated with covariates and an intercept in the measurement equation are easier to separate than process autocorrelation (equivalent to $\rho$), measurement error and process stochasticity. Note, however, that all of these parameters had cases associated with estimation problems. For example, the coefficient estimates were biased when their true simulated value was not equal to zero. \citet{Humbert2009a} suggested that in the case of exponential growth SSMs the population trend parameter, similar to an intercept in the process equation, was often well estimated and that increasing the precision of the abundance estimates and the length of the time series, more than the completeness of the time series, could increase the performance of the SSM. This further indicates that ecologists should closely consider model formulation, and that the estimability of parameter should be assessed.

If we cannot resolve the parameter estimation problem, we need to account for its potential effect on our inference. One way to account for the estimation uncertainty is to use a parametric bootstrap to get CIs on the parameter and state estimates \citep{Dennis2006,Forester2007}. These bootstrap CIs require simulating the model using the estimated parameter values and re-fitting the model to each simulation. The 2.5\textsuperscript{th} and  97.5\textsuperscript{th} quantiles of the estimated parameters and states then becomes the 95\% CI. These CIs differ from those we calculated from the standard deviation reported by \texttt{TMB}. However, because \texttt{TMB} is orders of magnitude faster than MCMC methods \citep{Albertsen2015}, implementing these parametric bootstrap CIs would be computationally feasible, even for complex models. Note, however, that the variability in the estimates of our simulations suggests that these CIs would be large and would often approach the boundary of parameter space.

\section{Conclusion}

We demonstrated that even simple linear Gaussian SSMs can have parameter estimability problems and that these problems can affect our ecological interpretation. As parameter estimability problems have been observed in other hierarchical models and because the ratio of information content to model complexity is expected to decrease with increasing numbers of hierarchies \citep{Lele2009,Lele2010a}, it is likely that these problems could occur in more complex forms of SSMs. Estimating individual variance components is notoriously difficult. SSMs do not escape this difficulty. While estimability problems have been discussed in the context of a few specific population dynamics SSMs \citep[e.g.,][]{Dennis2006, Knape2008,Polansky2009}, the voluminous literature on SSMs has paid relatively little attention to these problems. Such limited appreciation of the estimation problem is particularly dangerous because SSMs are usually advertised as providing the means to differentiate process from measurement variability \citep[e.g.,][]{deValpine2002,Patterson2008,Ahrestani2013}.

It is timely to warn ecologists of these difficulties. SSMs are becoming the favoured framework for animal movement and population dynamics. SSMs used in ecology are becoming increasingly complex \citep[e.g.,][]{McClintock2012}. In addition, tools to apply SSMs to data are becoming increasingly available. For example, \texttt{R} now provides a variety of packages that fit SSMs \citep{Petris2011}. Until recently, SSMs were applied by statisticians or by ecologists with a strong statistical background. These researchers were more likely to be aware of potential estimability problems than most ecologists. Researchers have questioned whether ecologists have sufficient statistical training to properly implement hierarchical models and have suggested that universities should start including advanced courses in statistical modelling in their ecological programs \citep[e.g.,][]{Dennis2006,Cam2012}. If the limitation of SSMs are not emphasized, the better accessibility of tools to fit these increasingly complex models are likely to lead to many undiagnosed estimation problems and incorrect conclusions.

While SSMs are powerful tools, they can give misleading results if they are misused. We believe it is important for ecologists to be aware of the potential estimation problems of SSMs. Investigating the likelihood profile, incorporating information on measurement error, and accounting for estimability uncertainty are all good first steps.  However, we urge statisticians to develop further tools that can be used to diagnosed such problems and these should be readily available along with the tools to fit SSMs.

\bibliography{LGSSM}

\section{Acknowledgements}

We thank William Aeberhard, Devin Lyons, and Stephanie Peacock for their inputs. This study was supported by grants to the Ocean Tracking Network from the Natural Sciences and Engineering Research Council of Canada (Research Network Grant NETGP 375118-08) and the Canada Foundation for Innovation, as well as by the Aquarium du Qu\'ebec, ArcticNet, US Department of Interior Bureau of Ocean Energy Management, Canadian Association of Zoos and Aquariums, Canadian Wildlife Federation, Circumpolar/Boreal Alberta Research, Environment Canada, Hauser Bears, Natural Sciences and Engineering Research Council of Canada, Northern Scientific Training Program, Polar Continental Shelf Program, Polar Bears International, Quark Expeditions, and World Wildlife Fund (Canada \& International). I.D.J. acknowledges a Macquarie Vice-Chancellor's Innovation Fellowship and M.A.L. acknowledges the Canada Research Chairs program.

\section{Contributions}

M.A.-M., J.M.F., C.F., M.A.L., and I.D.J. conceived the analyses,  M.A.-M., and C.M.A. analysed the results. M.A.-M. and A.E.D. conducted the field work. M.A.-M. with the help of all co-authors wrote the manuscript. All authors reviewed the manuscript.

\section{Additional information}

\textbf{Competing financial interests}

The authors declare no competing financial interests.

\clearpage

\begin{figure}[!htbp]
	\includegraphics{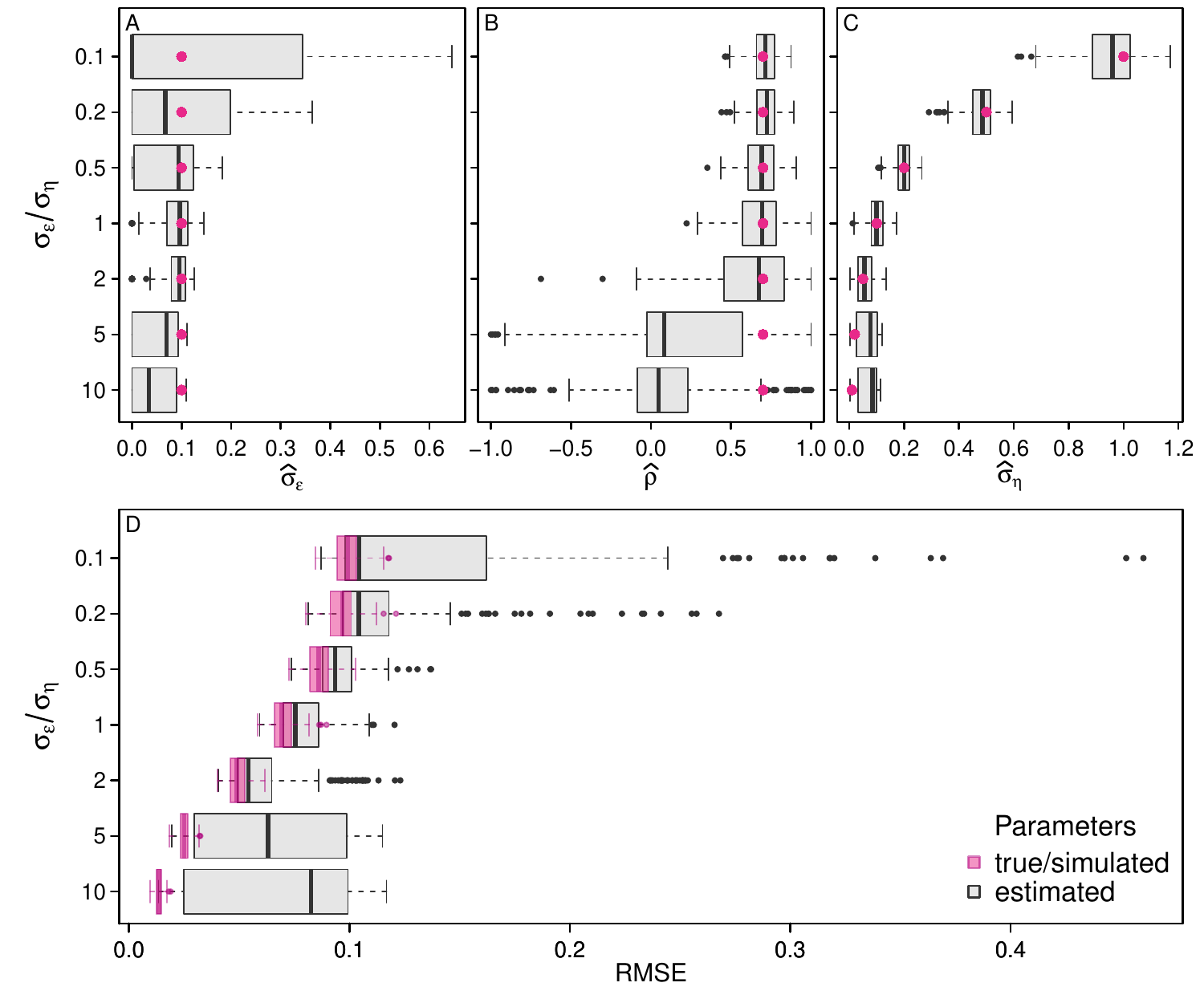}
	\caption{}
	\label{fig:TMBParEstSum}
\end{figure}

\textbf{Figure \ref{fig:TMBParEstSum}:} Changes in parameter estimates and state RMSEs associated with varying the measurement error to process stochasticity ratios ($\sigma_{\epsilon}/\sigma_{\eta}$) in the simulations. A-C) The boxplots represent the distribution of the parameter estimates ($\widehat{\sigma}_{\epsilon}$, $\widehat{\rho}$, $\widehat{\sigma}_{\eta}$) and the pink circles represent the true (simulated) values. D) The grey boxplots represent the distribution of the RMSE of the model fitted using the estimated parameter values, while the pink boxplots represent the RMSE when the model is fitted using the true parameter values.

\begin{figure}[!htbp]
	\includegraphics{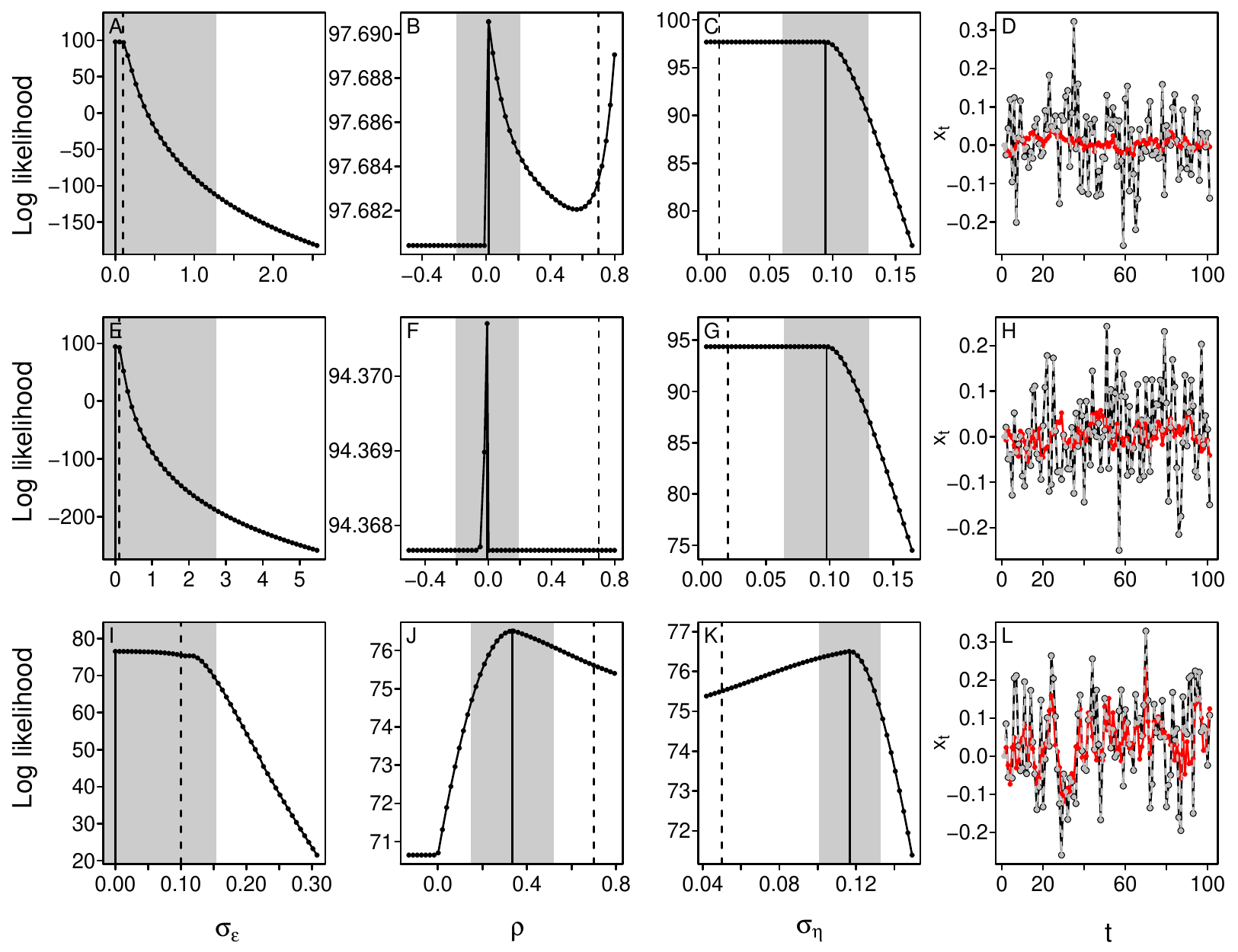}
	\caption{}
	\label{fig:LProfAll}
\end{figure}

\textbf{Figure \ref{fig:LProfAll}:} Log likelihood profiles for problematic simulations. In the first three columns, the curve represents the log likelihood when the focal parameter is fixed (the other parameter are optimise to maximise the log likelihood). The dash lines are the true parameter values (i.e., value used for the simulation), the full lines are the maximum likelihood estimates and the grey bands represent the $95\%$ CI. The last column shows the time-series. The black lines represent the observations, $y_t$, the red lines the simulated true states, $x_t$, and the grey dashed lines the estimated states, $\hat{x}_t$.

\begin{figure}[!htbp]
	\includegraphics{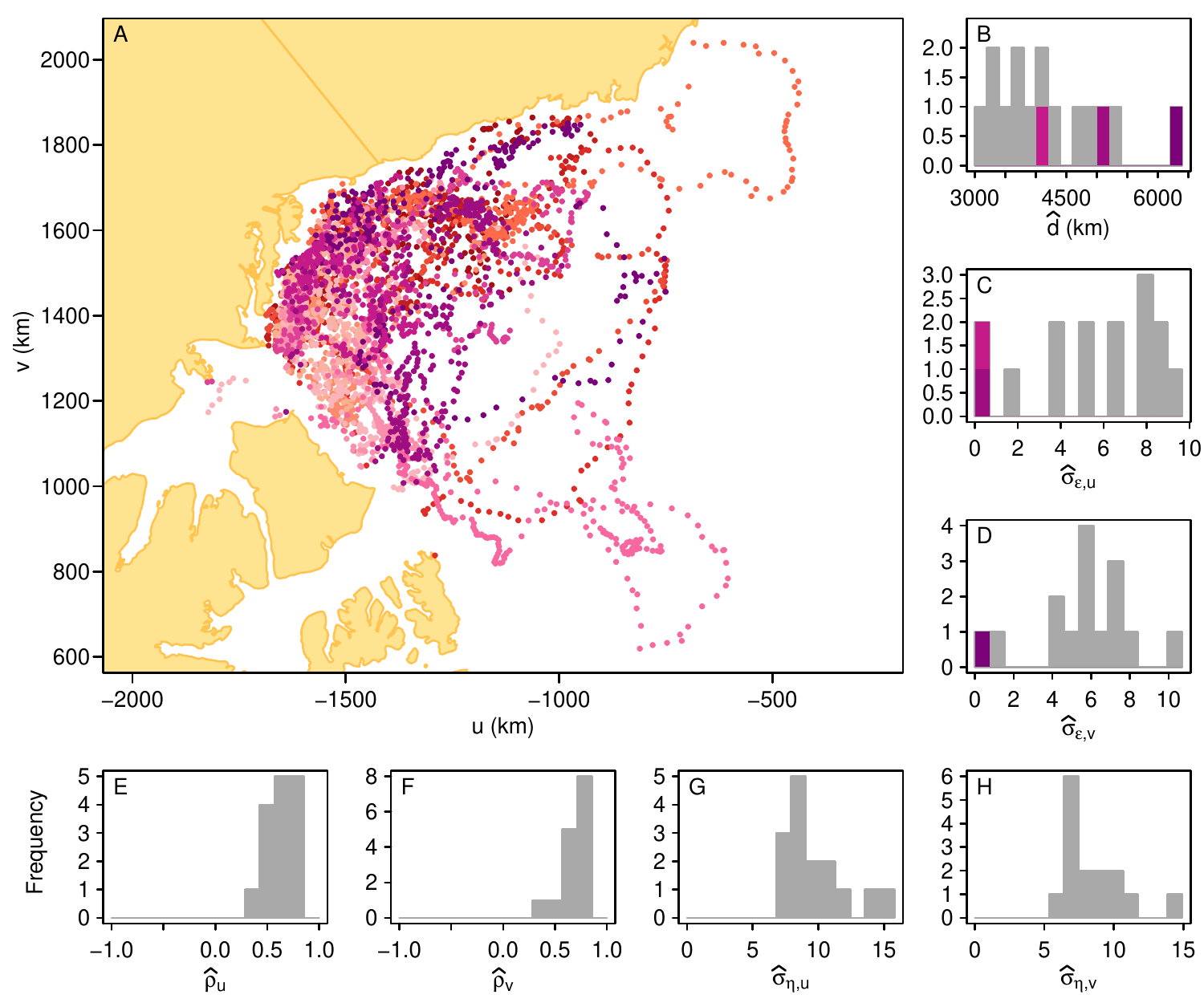}
	\caption{}
	\label{fig:Bear}
\end{figure}

\textbf{Figure \ref{fig:Bear}:} Polar bear movement, parameter estimates of the polar bear sea ice model, and estimates of the total voluntary bear displacement. A) Locations of the 15 polar bears used in the analysis, with colours representing different individuals. The map was created in \texttt{R} \citep{R2015} using the Northern Hemisphere azimuthal equal-area EASE-Grid projection developed for polar sea ice data  \cite{Brodzik2002}. B) Estimated total voluntary displacement over 342 days. C-H) Parameter estimates of the polar bear sea ice models. The different colours in panels B-C represent the three individuals for which either $\widehat{\sigma}_{\epsilon,u} < 0.01$ or $\widehat{\sigma}_{\epsilon,v} < 0.01$.

\clearpage

\appendix
\renewcommand\thefigure{\thesection.\arabic{figure}}
\renewcommand\thetable{\thesection.\arabic{table}}    

\FloatBarrier
\section{Some problems persist when $\rho$ is close to 1 and in the local-level model}
\label{App:Rho1}
\setcounter{figure}{0} 

\citet{Knape2008}, who investigated a linear Gaussian SSM describing the stochastic Gompertz population model, found that an autocorrelation parameter similar to $\rho$ was harder to estimate when its true simulated value was close to 0. $\rho$ appears to be problematic especially when $\lvert\rho\rvert < 0.5$. Similarly, \citet{Forester2007}, who developed a linear Gaussian SSM for animal movement, noticed that when the autocorrelation was close to 1 (i.e., 0.95) there was less estimation problems than when it was close to 0 (i.e., 0 or 0.2). This is not surprising. As the process becomes less autocorrelated it is harder to differentiate it from the temporally independent measurement error. As such, we focussed on investigating whether the estimation problems remained when the autocorrelation parameter was relatively high. In the main text, we have presented the results when $\rho = 0.7$. In this Appendix, we investigated higher $\rho$ values. First, we recreated the same simulation study as in section \ref{sec:sim} of the main text, except that we used $\rho=0.99$ in the simulations. Second, we used a simpler model called the local model, which is sometimes referred as the random walk plus noise \citep[e.g.,][]{Petris2010}:
\begin{linenomath}
\begin{align}
	\text{Measurement eq} && y_t &= x_t + \epsilon_t \,, & \epsilon_t \sim N(0, \sigma_{\epsilon}^2) \, & \text{, where } \, t \geq 1, \sigma_{\epsilon}^2 >0 \label{eq:mesEqLM}\\
	\text{Process eq} && x_t &=  x_{t-1} + \eta_t \,, & \eta_t \sim N(0, \sigma_{\eta}^2) \, & \text{, where } \, t \geq 1, \sigma_{\eta}^2 >0. \label{eq:procEqLM}
\end{align}
\end{linenomath}
The only difference with the model presented in eq. \ref{eq:mesEq}-\ref{eq:procEq} is that there is no $\rho$ parameters, which is the equivalent of fixing $\rho=1$. Note that while this simpler model has fewer parameters to estimate, it is no longer stationary \citep{Durbin2001}. Following the methods described in section \ref{sec:sim}, we simulated and fitted this simpler model. 

\begin{figure}[!htbp]
	\includegraphics{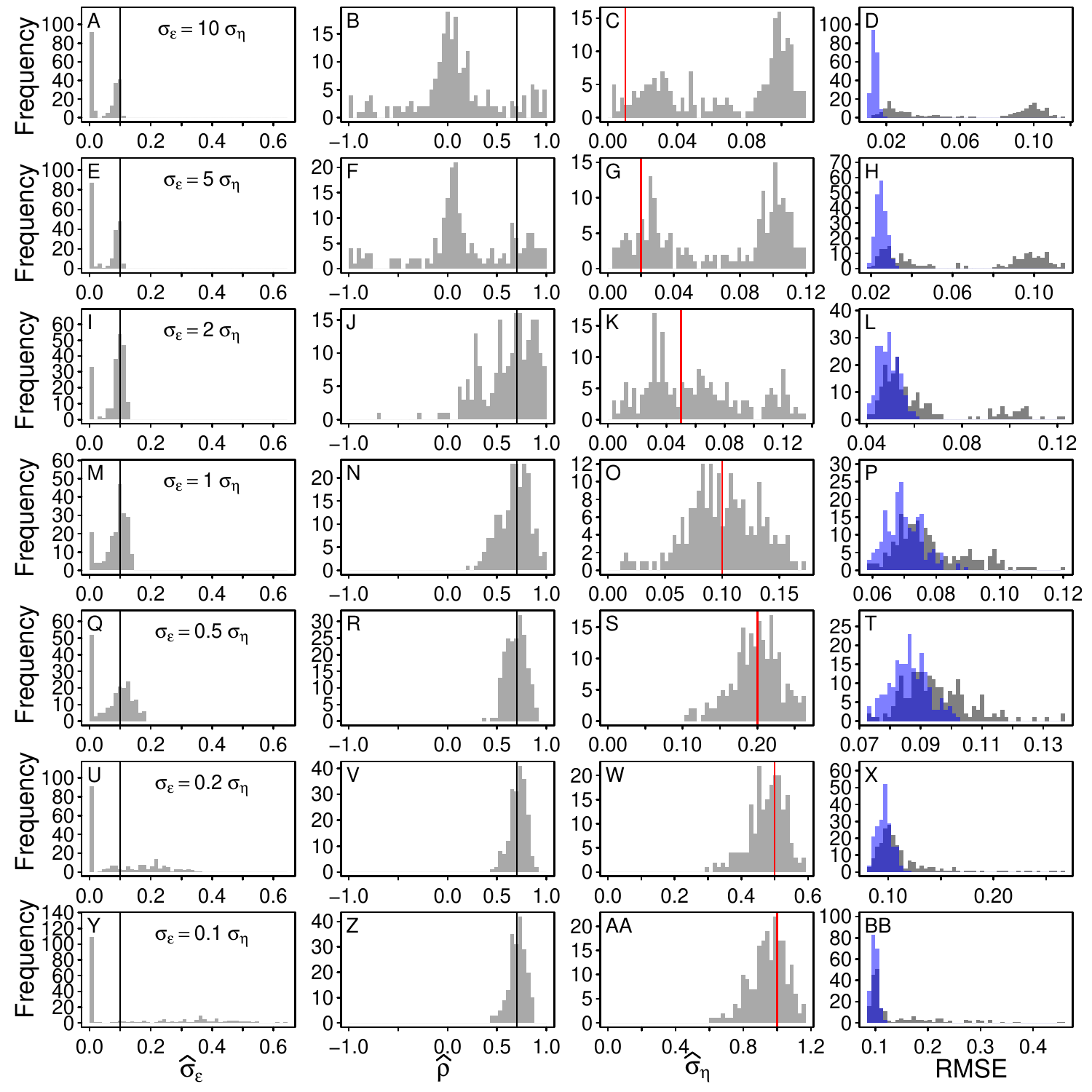}
	\caption{Histograms of the parameter estimates and of the RMSE of the state estimates when $\rho=0.7$. This is a more detailed visualization of the results presented in Fig. \ref{fig:TMBParEstSum} of the main text. Each row represents the results of 200 simulations for a set of parameter values. For the first three columns, the vertical lines represent the parameter values used in the simulations, with black lines used for values that remained constant, $\sigma_{\epsilon}=0.1$ and $\rho=0.7$, and red lines for values that changed between sets, $\sigma_{\eta} = (0.01,0.02,0.05,0.1,0.2,0.5,1)$. In the last column, the grey histograms represent the RMSE of the model fitted using the estimated parameter values, while the blue histograms represent the RMSE when the model is fitted using the true values.}
	\label{fig:TMBParEst}
\end{figure}

\begin{figure}[!htbp]
	\includegraphics{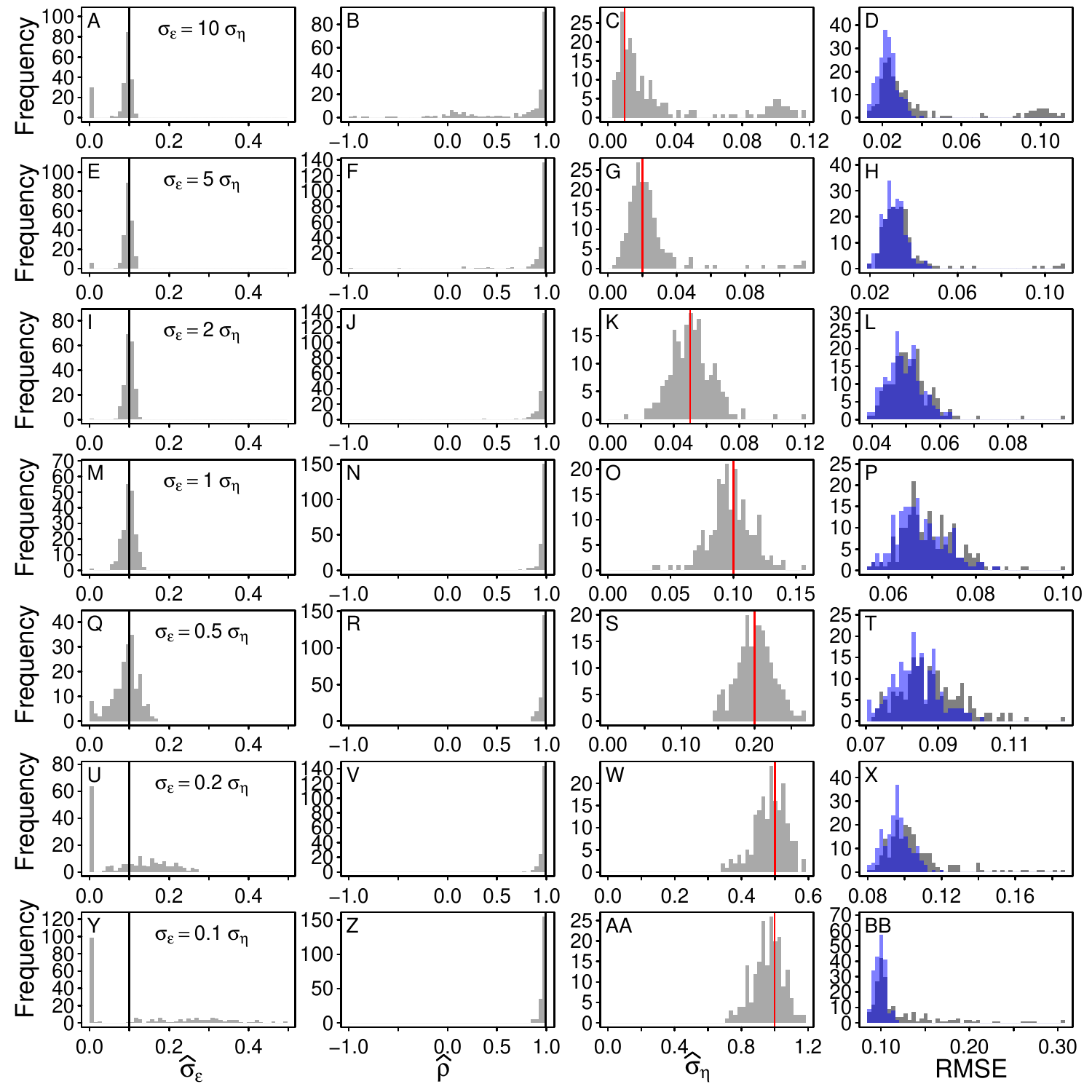}
	\caption{Histograms of the parameter estimates and of the RMSE of the states when $\rho=0.99$. Each row represents the results of 200 simulations for a set of parameter values. For the first three columns, the vertical lines represent the parameter values used in the simulations, with black lines used for the values that remain constant for all simulation sets, $\sigma_{\epsilon}=0.1$ and $\rho=0.99$, and red lines for values that change between set, $\sigma_{\eta} = (0.01,0.02,0.05,0.1,0.2,0.5,1)$. In the last column, the grey histograms represent the RMSE of the model fitted using the estimated parameter values, while the blue histograms represent the RMSE when the model is fitted using the simulation values.}
	\label{fig:TMBParEst99}
\end{figure}

\begin{figure}[!htbp]
	\includegraphics{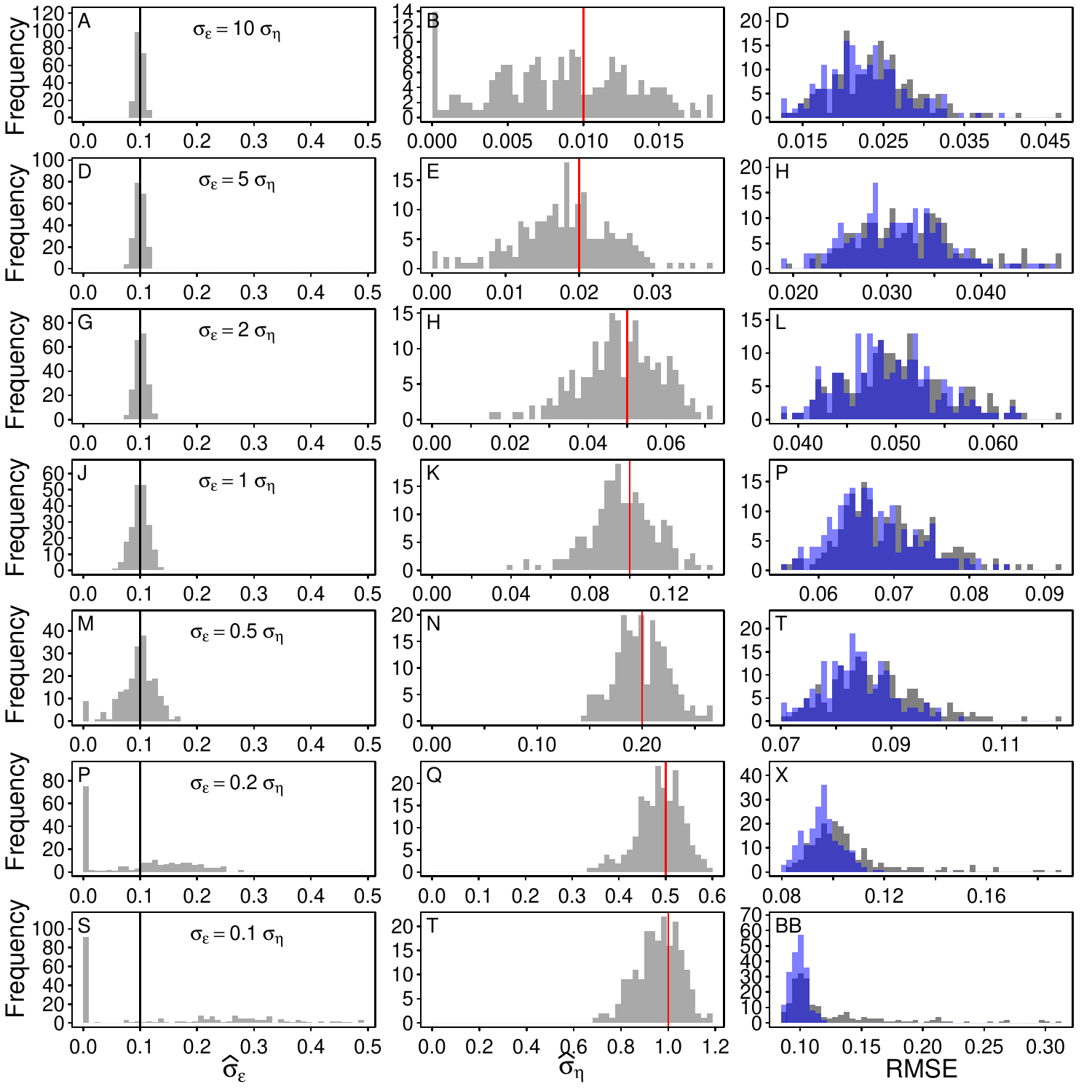}
	\caption{Histograms of the parameter estimates and of the RMSE of the states for a set of simulations of the local-level model. Each row represents the results of 200 simulations for a set of parameter values. For the first three columns, the vertical lines represent the parameter values used in the simulations, with black lines used for the values that remain constant for all simulation sets, $\sigma_{\epsilon}=0.1$, and red lines for values that change between set, $\sigma_{\eta} = (0.01,0.02,0.05,0.1,0.2,0.5,1)$. In the last column the grey histograms represent the RMSE of the model fitted using the estimated parameter values, while the blue histograms represent the RMSE when the model is fitted using the simulation values.}
	\label{fig:TMBParEstLM}
\end{figure}

The parameter and state estimates improved in simulations with $\rho=0.99$ (Fig. \ref{fig:TMBParEst99}). In particular, the estimates for the process parameters ($\rho$, $\sigma_{\eta}$) were much closer to their simulated values than when the simulated $\rho$ value was 0.7 (e.g., compare Fig. \ref{fig:TMBParEst99}B-C to Fig. \ref{fig:TMBParEst}B-C). In some cases, this translated in better state estimates (e.g., compare Fig. \ref{fig:TMBParEst99}H to Fig. \ref{fig:TMBParEst}H). Similarly, using the local model improved parameter and states estimates (Fig. \ref{fig:TMBParEstLM}). However, this did not completely eliminate the estimation problems. When the measurement error was much larger than the process stochasticity, $\sigma_{\epsilon} = 10 \; \sigma_{\eta}$, some of the parameter estimates remained on the boundary of parameter space (e.g., Fig. \ref{fig:TMBParEst99}A and Fig. \ref{fig:TMBParEstLM}B). In the case of simulations with $\rho =0.99$, the estimated $\rho$ was close to 0 (Fig. \ref{fig:TMBParEst99}B,F) and some state estimates were far from those estimated when the parameter values were known (e.g, Fig. \ref{fig:TMBParEst99}D,H).

\FloatBarrier
\section{Some problems persist with longer time-series}
\label{App:n500}

Using longer time-series can considerably reduce estimability problems. In this appendix, we investigate whether the estimation problem persisted with longer time-series. We wanted to use a length of time-series that is relevant to real ecological examples, keeping in mind that, generally, population abundance time-series are shorter than movement time-series. For example, a recent population study using bird count data was limited to 45 to 52 counts \citep{Simmons2015}. As such previous simulation studies for population dynamics SSMs limited their time-series to 30 and 100 time steps \citep{Dennis2006, Knape2008,Humbert2009a}. In fact, \citet{Dennis2006} mentioned that times-series of 100 time steps were unrealistic for ecological data of the type and thus our example in the main text ($n=100$) is likely underestimating the frequency of estimation problems. Movement time-series are generally of longer length. For example, the simulation study of \citet{Forester2007} used 350 steps and their real movement data ranged from 265-390 locations. Our polar bear time-series ranged from 342 to 365. In addition, with technological advancements movement time-series are becoming much longer. Thus, to look at time-series more representative of movement time-series we conducted the same simulation analysis as in section \ref{sec:sim} of the main text, with the only difference being that our time-series have 500 observations ($n=500$) rather than 100 ($n=100$). 

When time-series of 500 steps were used, the estimation problems were reduced (compare Fig. \ref{fig:TMBParEstn500} to Fig. \ref{fig:TMBParEst}). In particular, when $\sigma_{\epsilon} \leq 2 \; \sigma_{\eta}$, we had fewer $\sigma_{\epsilon}$ estimates at the boundary of parameter space (e.g., compare Fig. \ref{fig:TMBParEstn500}Q to Fig. \ref{fig:TMBParEst}Q) and the estimates of $\sigma_{\eta}$ and $\rho$ are closer to the simulated values (e.g., compare Fig. \ref{fig:TMBParEstn500}N,O to Fig. \ref{fig:TMBParEst}N,O). However, when the measurement error is large compared to the process stochasticity, $\sigma_{\epsilon} \geq 5 \; \sigma_{\eta}$, many $\sigma_{\epsilon}$ estimate remained at the boundary of parameter space (Fig. \ref{fig:TMBParEstn500}A,E), many  $\sigma_{\eta}$ estimates were positively biased (Fig. \ref{fig:TMBParEstn500}C,G), and many $\rho$ estimates were negatively biased, with values close to 0 (Fig. \ref{fig:TMBParEstn500}B,F). In addition, when the parameters were estimated, many replicates had higher state estimate error than when the true parameter values were used (Fig. \ref{fig:TMBParEstn500}D,H), indicating that biases in parameter estimates continued to affect the state estimates. Overall, these results suggest that longer time-series do improve the estimability of some parameters and states, but that to have reliable estimates when the measurement error is much larger than the process stochasticity would require much longer time-series.

\setcounter{figure}{0} 
\begin{figure}[!htbp]
	\includegraphics{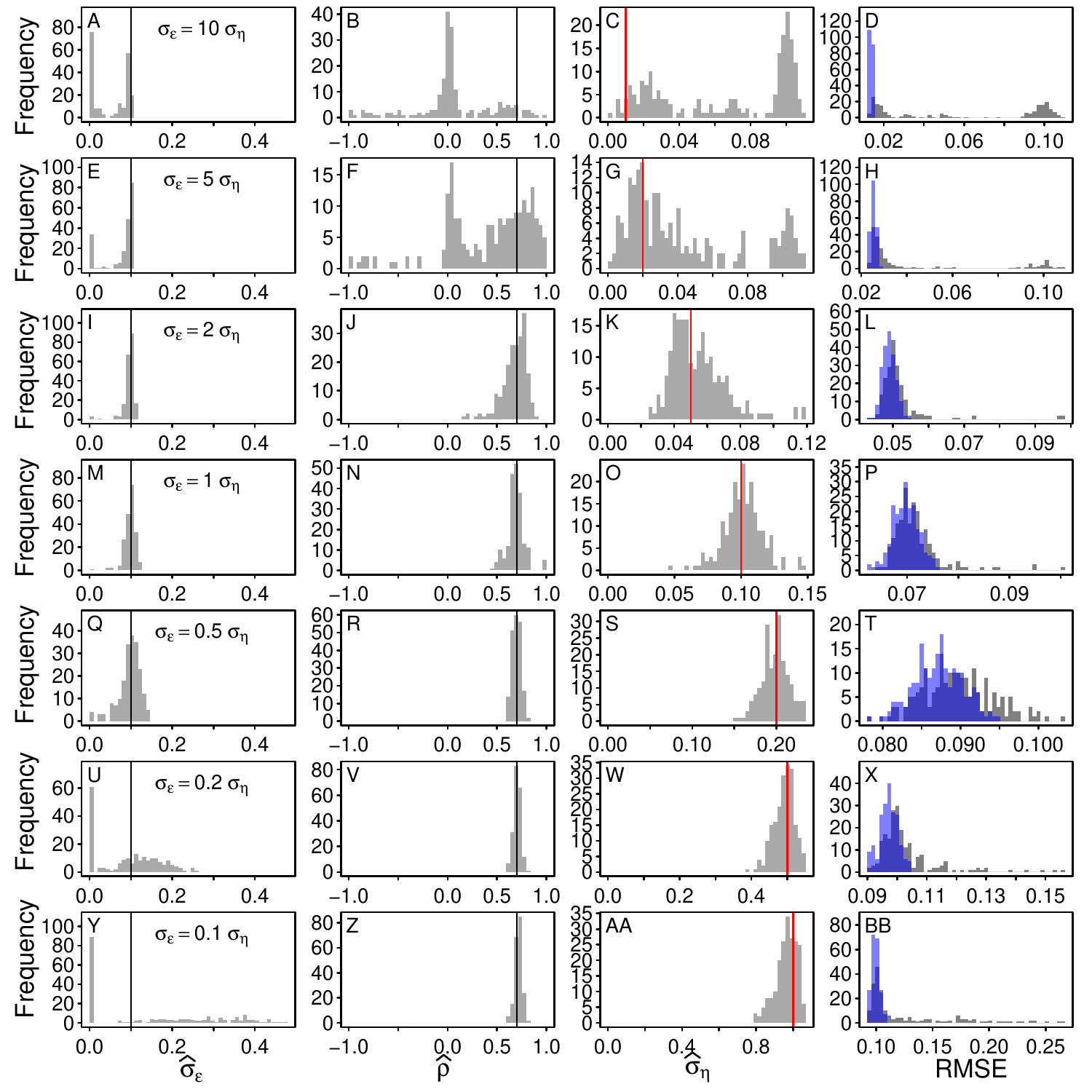}
	\caption{Histograms of the parameter estimates and of the RMSE of the states for a set of simulations with 500 time steps. Each row represents the results of 200 simulations for a set of parameter values. For the first three columns, the vertical lines represent the parameter values used in the simulations, with black lines used for the values that remain constant, $\sigma_{\epsilon}=0.1$ and $\rho=0.7$, and red lines for values that change between set, $\sigma_{\eta} = (0.01,0.02,0.05,0.1,0.2,0.5,1)$. In the last column, the grey histograms represent the RMSE of the model fitted using the estimated parameter values, while the blue histograms represent the RMSE when the model is fitted using the simulation values.}
	\label{fig:TMBParEstn500}
\end{figure}

\FloatBarrier
\setcounter{figure}{0} 
\section{Results with other R packages}
\label{App:DLMandJAGS}

We chose to use \texttt{TMB} for most analyses because it is a fast and flexible package that can be used to fit a variety SSMs to data \citep[e.g.,][]{Albertsen2015}. To verify that the estimation problems are not limited to this package and are general problems associated with linear Gaussian SSMs, we also used two additional packages to reproduce the simulation study explained in section \ref{sec:sim} of the main text. First, we used \texttt{dlm} \citep{Petris2010}, a package that was used in recent ecological studies to fit SSMs to data \citep{Tittensor2014a,Simmons2015}. The package \texttt{dlm} maximises the log likelihood numerically and has functions to estimate the states via Kalman filter and smoother. To be consistent with TMB, we used the Kalman smoother, which takes into account all observations \citep[see][]{Albertsen2015}. 

Second, we used \texttt{rjags} \citep{Plummer2014}, which is an \texttt{R} interface to JAGS, a program that allows for the analysis of Bayesian hierarchical models using Markov Chain Monte Carlo (MCMC) methods. Unlike \texttt{TMB} and \texttt{dlm}, \texttt{rjags} requires the specification of priors for the estimated parameter. We used the vague priors: $\sigma_{\epsilon} \sim \text{HalfN}(0, \sigma^2=10000)$, $\rho \sim \text{Uniform}(0, 1)$, and $\sigma_{\eta} \sim \text{HalfN}(0, \sigma^2=10000)$. We used two chains, each with 50 000 adaptation steps, 50 000 burn in steps and 50 000 saved steps. For each chain we kept 1 every 500 steps.

\begin{figure}[!htbp]
	\includegraphics{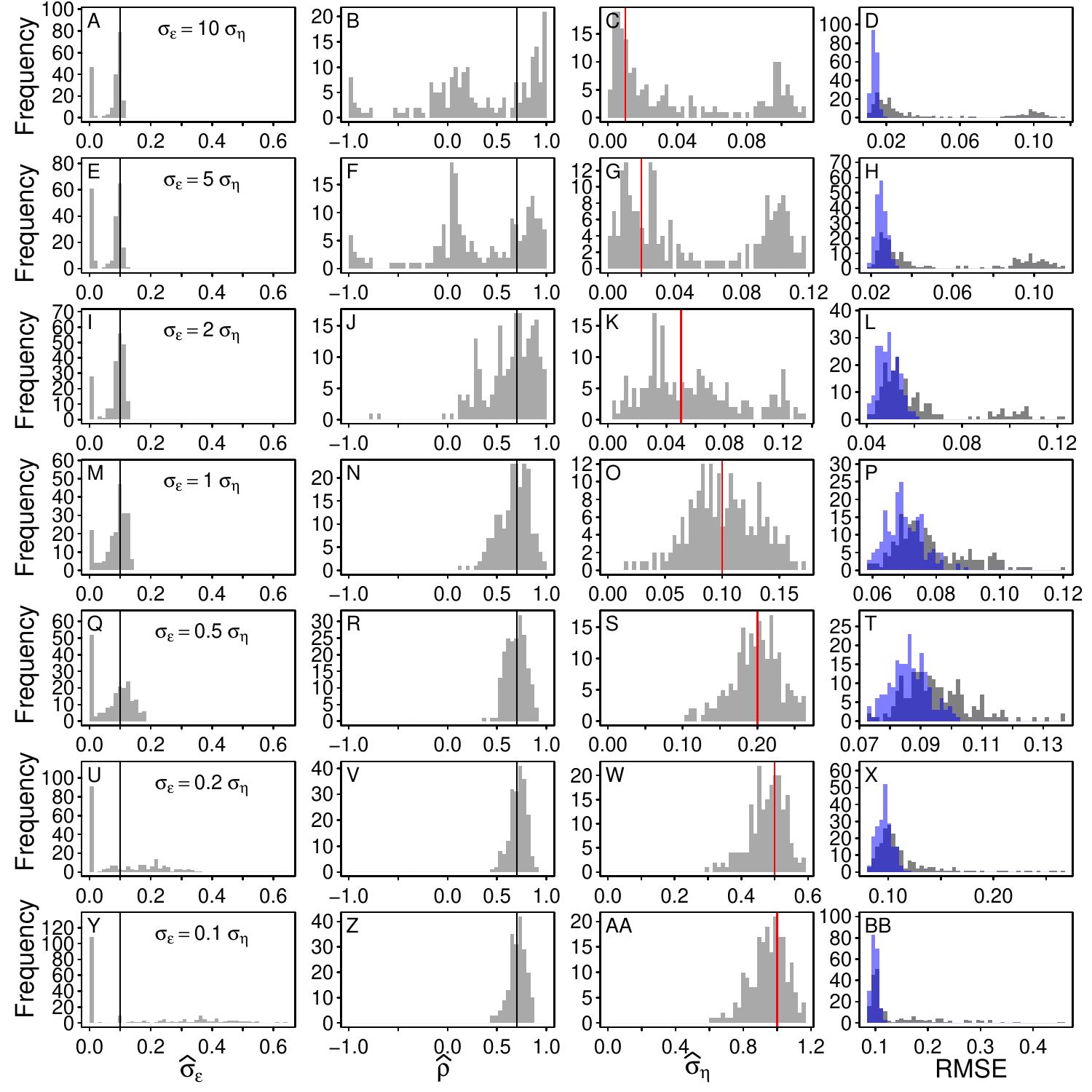}
	\caption{Histograms of the parameter estimates and of the RMSE of the states when \texttt{dlm} is used to fit our SSM to a set of simulations. Each row represents the results of 200 simulations for a set of parameter values. For the first three columns, the vertical lines represent the parameter values used in the simulations, with the black lines used for the values that remain constant, $\sigma_{\epsilon}=0.1$ and $\rho=0.7$, and the red lines for values that change between set, $\sigma_{\eta} = (0.01,0.02,0.05,0.1,0.2,0.5,1)$. In the last column, the grey histograms represent the RMSE of the model fitted using the estimated parameter values, while the blue histograms represent the RMSE when the model is fitted using the simulation values. Note that the parameters of 2 out of the 1400 simulations could not be estimated due to singularity problems.}
	\label{fig:dlmParEst}
\end{figure}

The results when we used \texttt{dlm} are nearly identical to those when we used \texttt{TMB} (compare Fig. \ref{fig:dlmParEst} to Fig. \ref{fig:TMBParEst}). The only difference, is that when $\sigma_{\epsilon} \geq 5 \; \sigma_{\eta}$ a few less replicates had $\sigma_{\epsilon}$ values close to 0, $\rho$ values close to 0, and positively biased $\sigma_{\eta}$ values. However, these differences were small and some were accompanied by other biases, such as more $\rho$ values close to -1. Overall, the conclusion made for \texttt{TMB} in the main text hold true for \texttt{dlm}.

In contrast, the results from \texttt{rjags} are different from \texttt{TMB}. In particular, the distribution of estimates were unimodal and there was no estimates close to the boundary of parameter space (i.e., no $\widehat{\sigma}_{\epsilon}$ close to 0). However, the peak of estimates was often far from the simulated value (Fig. \ref{fig:jagsParEst}), indicating that both parameter and state estimates were often biased. This is potentially due to the fact that vague priors can influence the results and smooth the peculiarities of the likelihood \citep{Dennis2006}. This effect could explain why estimation problems have been less easily detected in recent SSMs studies, which often uses complex Bayesian SSMs. In addition, the posterior distributions were more unimodal and not as flat as the likelihood profiles produced by likelihood-based methods (compare Fig. \ref{fig:jagsParEst} to Fig. \ref{fig:LProfAll}). Note that these results exclude the 37 replicates out of 1400 simulations that did not converge (scale reduction factor of any parameter $>1.1$), and thus would have been deemed problematic with such metrics. Overall, the results from JAGS indicate that using Bayesian methods does not fix the estimation problems of the linear Gaussian SSMs, and in fact might have made them harder to detect. See Appendix \ref{App:diagTools} for a more detailed discussion of diagnostic tools.

\begin{figure}[!htbp]
	\includegraphics{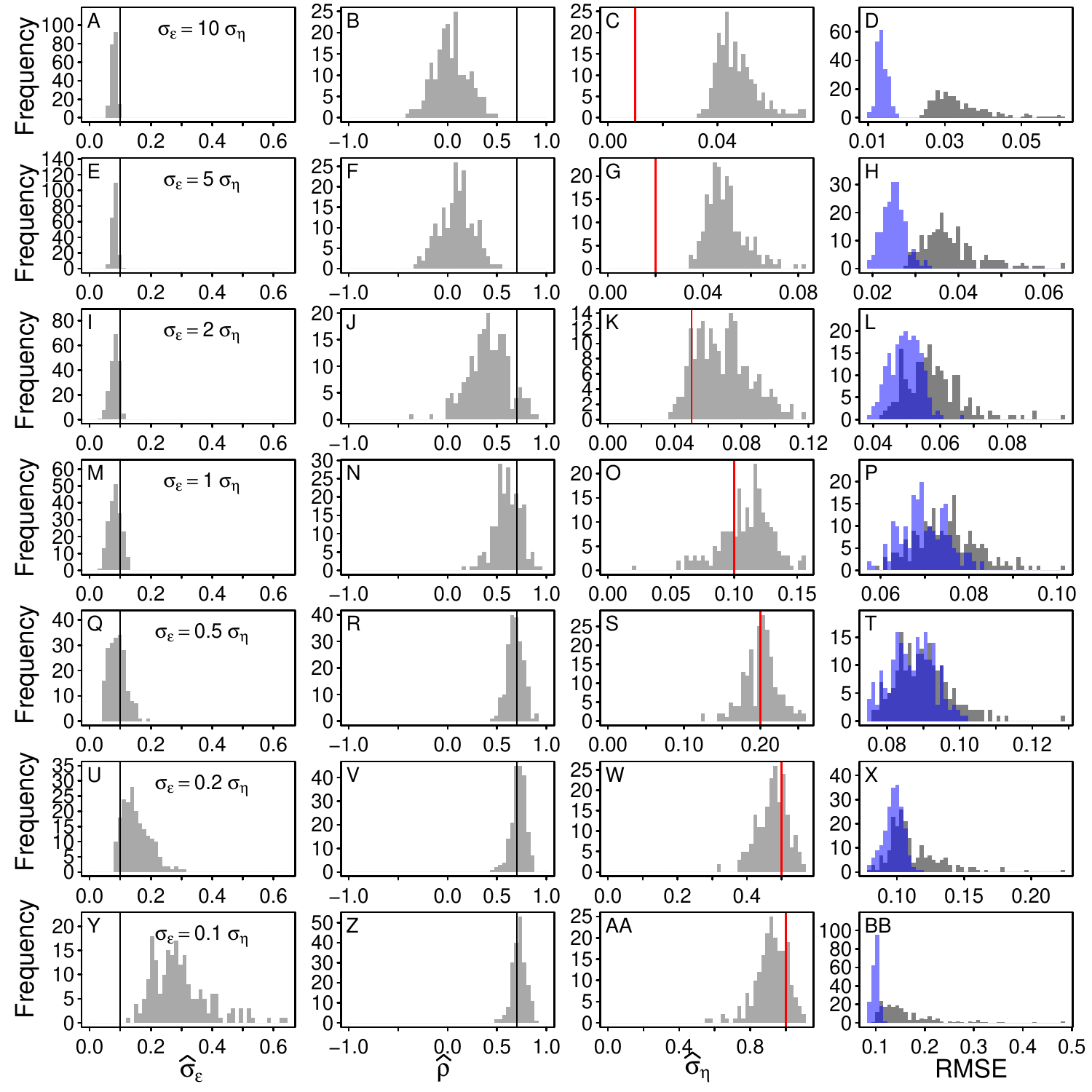}
	\caption{Histograms of the parameter estimates and of the RMSE of the states when \texttt{rjags} is used to fit our SSM to a set of simulations. Each row represents the results of 200 simulations for a set of parameter values. For the first three columns, the vertical lines represent the parameter values used in the simulations. The black lines are used for the values that remain constant for all simulation sets: $\sigma_{\epsilon}=0.1$ and $\rho=0.7$. The red line for that values that change in between set: $\sigma_{\eta} = (0.01,0.02,0.05,0.1,0.2,0.5,1)$. In the last column the grey histograms represent the RMSE of the model fitted using the estimated parameter values, while the blue histograms represent the RMSE when the model is fitted using the simulation values. Note that 37 out of 1400 simulations did not converged (i.e. the potential scale reduction factor was $> 1.1$ for one of the parameters).}
	\label{fig:jagsParEst}
\end{figure}

\begin{figure}[!htbp]
	\includegraphics{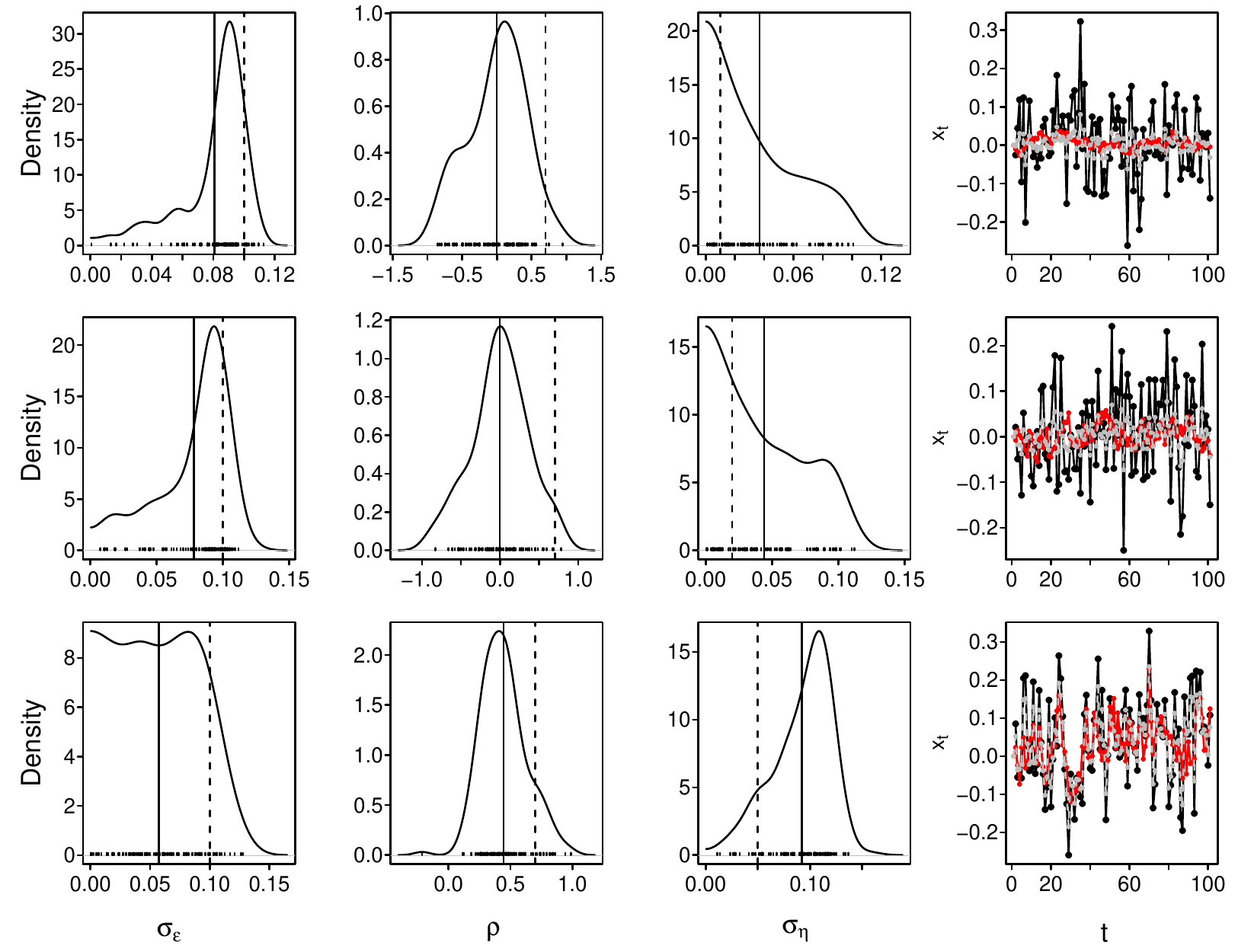}
	\caption{Posterior distribution for problematic simulations. The first three columns, the curve represents the posterior distribution for the estimated parameters. The dash lines are the true parameter values (i.e., value used for the simulation), the full lines are the mean value. The last column shows the time-series. The black lines represent the observations, $y_t$, the red lines the simulated true states, $x_t$, and the grey dashed lines the estimated states, $\hat{x}_t$. Note that these three simulations converged (the potential scale reduction factor was $< 1.1$ for all parameters).}
	\label{fig:jagsParEst}
\end{figure}

\FloatBarrier
\setcounter{figure}{0} 
\setcounter{table}{0} 
\section{Estimating parameters when we know the true states}
\label{App:FixX}

\begin{figure}[!htbp]
	\includegraphics{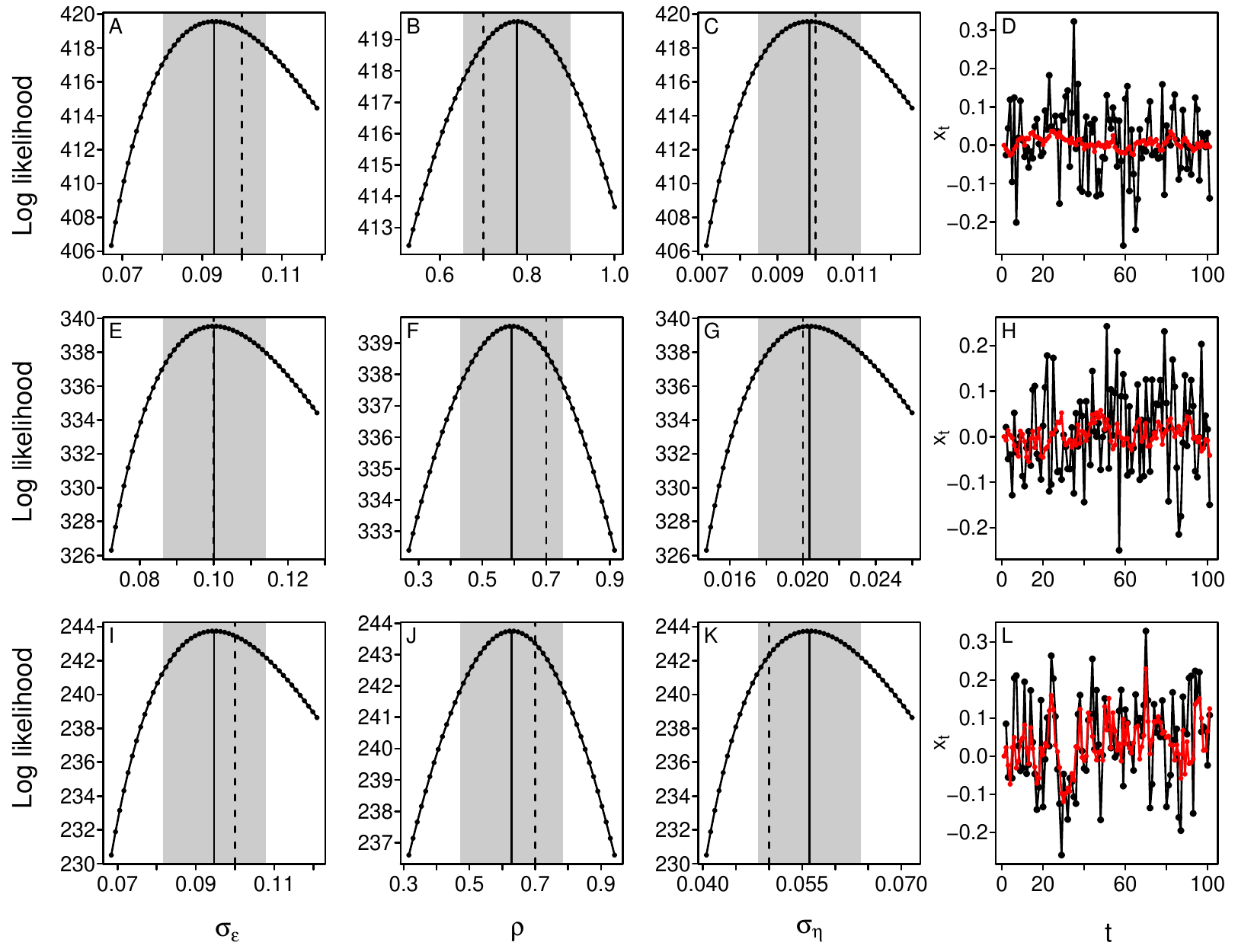}
	\caption{Log likelihood profile for problematic simulations when the state values are fixed to the simulated values. In the first three columns, the curve represents the log likelihood when the focal parameter is fixed (the other parameter are optimize to maximize the log likelihood). The dash lines are the true parameter values (i.e., value used for the simulation), the full lines are the maximum likelihood estimates and the grey bands represent the $95\%$ CIs. The last column shows the time-series. The black lines represent the observations, $y_t$, and the red lines the simulated true states, $x_t$. Note that these are the same simulations as in Fig. \ref{fig:LProfAll}.}
	\label{fig:LProffixx}
\end{figure}

In this Appendix we investigate whether the parameter-estimation problems are associated with estimating the parameters at the same time as the states. To do so, we estimated the parameters when the states were fixed to their true simulated values. We focused on the same problematic simulations as those explored in the main text. When the states were known, the parameter estimates were close to the simulated values and the CIs included the true simulated values (Fig. \ref{fig:LProffixx}). In addition, the likelihood profiles were unimodal, demonstrating the type of likelihood profiles one would expect for well-behaved models. These results suggest that the problems lie in estimating both the states and the parameters.

\FloatBarrier
\setcounter{figure}{0} 
\setcounter{table}{0} 
\section{Reformulating our simple SSMs with the ARMA notation}
\label{App:ARMA}

For certain parameter values, the SSM can behave either as a white noise process or as an AR(1) process. In both cases, the SSM formulation of the model will be over-parameterized, and will lead to estimation problems.
To see this, we can rewrite our SSM as an ARMA(1,1) model. First, we can combine eq. \ref{eq:mesEq} and \ref{eq:procEq} and reparametrize the model in terms of $\epsilon_t$:
\begin{linenomath}
\begin{align}
	y_t &= \rho x_{t-1} + \eta_t + \epsilon_t , & \epsilon_t \sim N(0,\sigma^2_{\epsilon}), \eta_t \sim N(0,\sigma^2_{\eta})\\
	&= \rho (x_{t-1}  + y_{t-1} - y_{t-1}) + \eta_t +\epsilon_t\\
	&= \rho y_{t-1} + \rho \epsilon_{t-1} + \eta_t +\epsilon_t
\end{align}
\end{linenomath}
Since $\eta_t$ and $\epsilon_t$ are independent and normally distributed, their sum, $\nu_t = \eta_t + \epsilon_t$, follows a normal distribution with zero mean and variance $\sigma^2_{\nu} = \sigma^2_{\eta} + \sigma^2_{\epsilon}$. Now, if we let $\nu_{t-1} \sim N(0,\sigma^2_{\nu})$, we can rescale its variance such that:
\begin{linenomath}
\begin{align}
	y_t &\overset{D}{=} \rho y_{t-1} + \rho \frac{\sigma_{\epsilon}}{\sqrt{\sigma^2_{\epsilon} + \sigma^2_{\eta}}} \nu_{t-1} +\nu_t.
\end{align}
\end{linenomath}
This is an ARMA(1,1) process with AR parameter $\phi= \rho$, MA parameter $\psi=  \rho \frac{\sigma_{\epsilon}}{\sqrt{\sigma^2_{\epsilon} + \sigma^2_{\eta}}}$, and variance parameter $\sigma^2_{\nu}$.

When $\sigma_{\epsilon} \ll \sigma_{\eta}$, then $\psi$ is small. Thus the process behaves as an AR(1) process with parameters $\rho$ and $\sigma_{\eta}$. This is the case for our simulations with $\sigma_{\eta} = 0.5, 1$ (Fig. \ref{fig:TMBParEst}). When $\sigma_{\epsilon} \gg \sigma_{\eta}$, then $\psi \approx \phi$ and hence there is parameter redundancy in the model \citep{Box2008}. In this case, the process closely resembles white noise. This is the case for our most problematic simulations ($\sigma_{\eta} = 0.01, 0.02, 0.05$, see Figs.  \ref{fig:TMBParEst}-\ref{fig:LProfAll}).

\FloatBarrier
\newpage
\setcounter{figure}{0} 
\setcounter{table}{0} 
\section{Polar bear and sea ice simulations}
\label{App:PBsim}

To demonstrate that the polar bear and sea ice model has estimation problems and show how these problems may affect our interpretation of our proxy of energy expenditure, we simulated movement using model described in eq. \ref{eq:pbX0}-\ref{eq:pbPro}. We simulated 500 movement paths with $n=342$ using the parameters estimated with the polar bear data (Table \ref{table:pbParEst}). We also used the initial state values estimated with the polar bear data: $\mathbf{x}_0 = \left[ \begin{smallmatrix} \;\:\:2.01 \\-4.08 \end{smallmatrix} \right]$. We simulated the sea ice displacement in the $u$- and $v$-direction using normal distributions with mean and standard deviation values based on the sea ice drift experienced by the polar bears in our sample: $s_{t,u} \sim N(\mu=1.49, \sigma=4.97)$ and $s_{t,v} \sim N(\mu=2.26, \sigma=3.97)$. As for the empirical data, we estimated parameters and the total voluntary displacement (see eq. \ref{eq:disp}). 

\begin{table}[!htbp]
	\caption{Parameter estimates for the polar bear sea ice empirical data. These are the parameter values used in the simulations.}
	\begin{tabular}{c c c}
		\hline
		Parameters & mean & sd \\
		\hline
		$\sigma_{\epsilon,u}$ & 5.53 & 3.10 \\
		$\sigma_{\epsilon,v}$ & 5.79 & 2.66 \\
		$\rho_{u}$ & 0.635 & 0.128 \\
		$\rho_{v}$ & 0.685 & 0.123 \\
		$\sigma_{\eta,u}$ & 9.66 & 2.47 \\
		$\sigma_{\eta,v}$ & 8.56 & 2.33 \\
		\hline
		\end{tabular}
	\label{table:pbParEst}
\end{table}

\begin{figure}[!htbp]
	\includegraphics{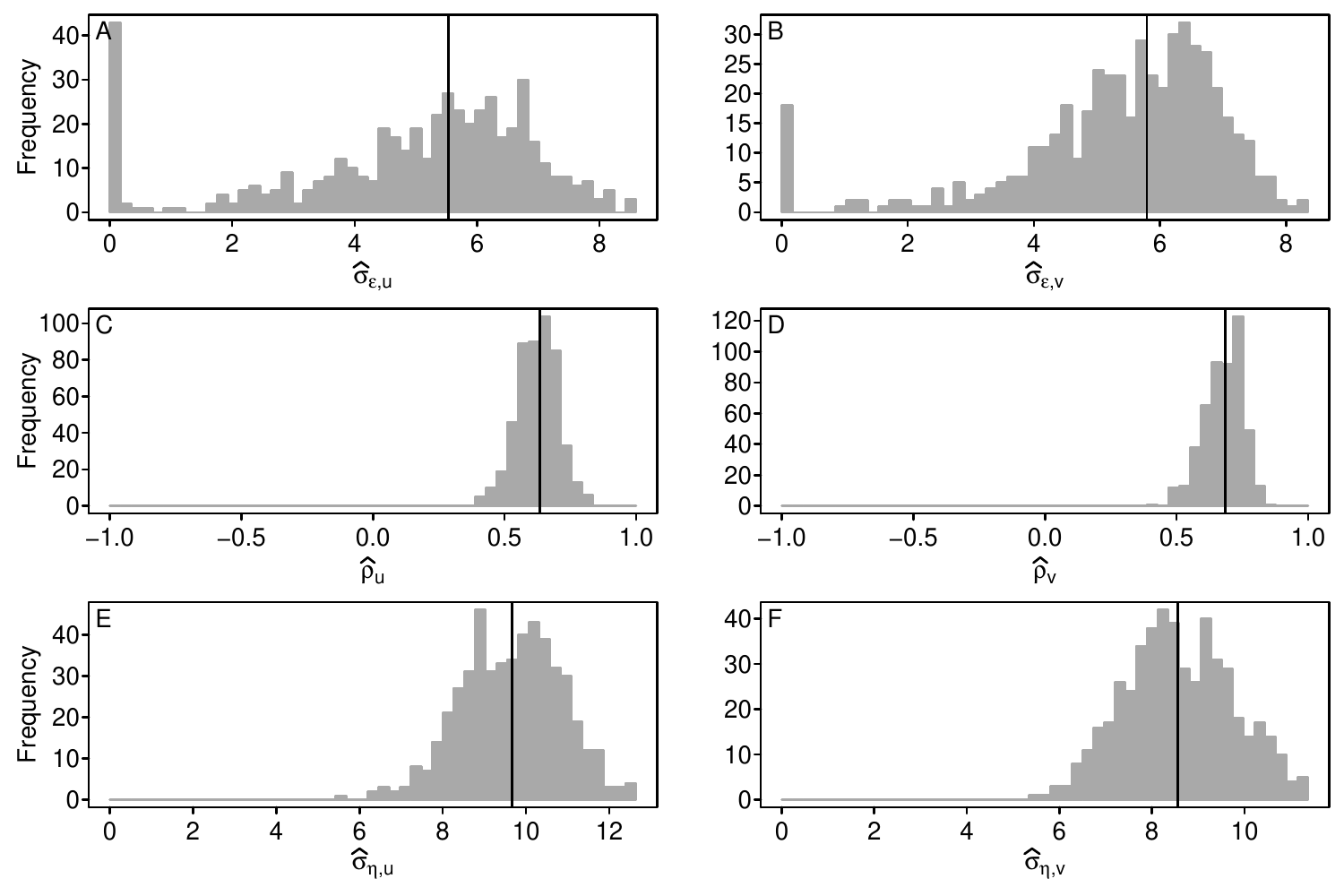}
	\caption{Histograms of the parameter estimates for the set of 500 simulations of the polar bear sea ice model. The vertical lines represent the parameter values used in the simulations. The black lines represent the value used to simulate the data (see Table \ref{table:pbParEst}).}
	\label{fig:pbSimParEst}
\end{figure}

\begin{figure}[!htbp]
	\includegraphics{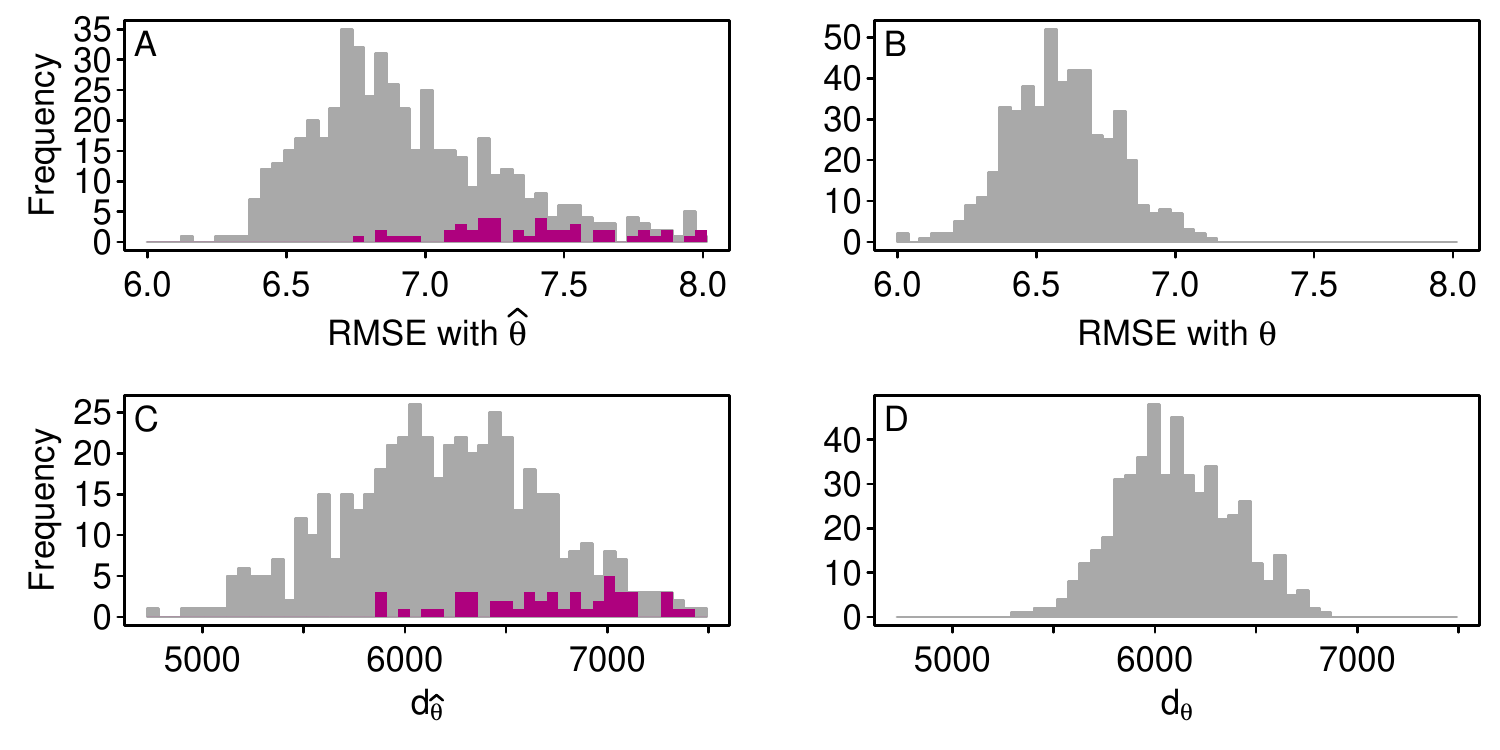}
	\caption{Histograms of the RMSE of the states for the polar bear sea ice model and of the total displacement, $d$. The left column represents results when the model was fitted using the estimated parameter values. The right column represents the results when the model was fitted using the simulation values. The purple columns represent the simulations for which either $\widehat{\sigma}_{\epsilon,u} < 0.01$ or $\widehat{\sigma}_{\epsilon,v} < 0.01$.}
	\label{fig:pbSimRMSE}
\end{figure}

Our simulation results show that the model appears to have similar estimation problems than the simple SSM extensively studied in the manuscript. In particular, we have a few simulations where the estimates of $\sigma_{\epsilon,u}$ and $\sigma_{\epsilon,v}$ are close to 0 (Fig. \ref{fig:pbSimParEst}), something that was also noticeable in the empirical data (Fig. \ref{fig:Bear}). Parameter estimates close to 0 can be associated with  estimation problems, which in turn can affect the state estimate and thus the estimates of the total displacement (Fig. \ref{fig:pbSimRMSE}). In particular, simulations with either $\widehat{\sigma}_{\epsilon,u} < 0.01$ or $\widehat{\sigma}_{\epsilon,v} < 0.01$ tended to be associated with higher $d$ values (Fig. \ref{fig:pbSimRMSE}). The estimates of $d$ ranged widely in values, but appeared to by generally higher.

\FloatBarrier
\clearpage
\setcounter{figure}{0} 
\setcounter{table}{0} 
\section{Results when the measurement error is fixed}
\label{App:mesFix}

As explained in section \ref{MesMethod} of the main text, we investigated whether fixing the measurement error resolved the parameter estimation problem. To do so, we fitted our simple likelihood (eqn. \ref{eq:lik}) to the same simulations as in section \ref{sec:sim}, but we fixed the standard deviation of the measurement equation to the value used to simulate the data, $\sigma_{\epsilon} = 0.1$. We only estimated the remaining parameters, $\mathbf{\theta_{m}} = (\rho, \sigma_{\eta})$. As for the main analysis, we investigated the parameter estimates, RMSE of the states, and likelihood profiles. In addition, we explored the likelihood surfaces.

\begin{figure}[!htbp]
	\includegraphics{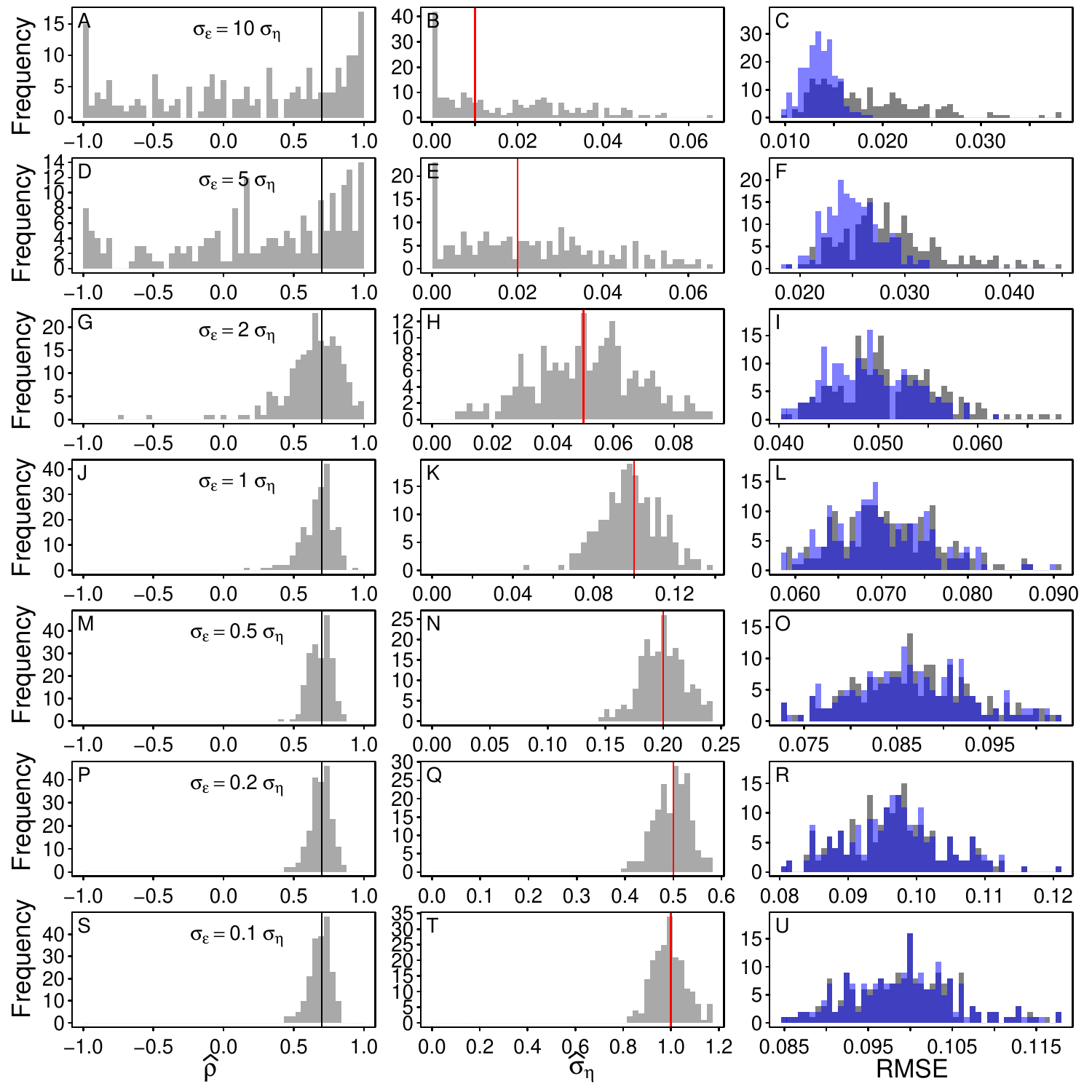}
	\caption{Histograms of the parameter estimates and of the RMSE of the states when the values of the measurement error is fixed to its true value. Each row represents the results of 200 simulations for a set of parameter values. For the first three columns, the vertical lines represent the parameter values used in the simulations, with black lines used for the values that remain constant, $\sigma_{\epsilon}=0.1$ and $\rho=0.7$, and the red lines for values that change between set, $\sigma_{\eta} = (0.01,0.02,0.05,0.1,0.2,0.5,1)$. In the last column, the grey histograms represent the RMSE of the model fitted using the estimated parameter values, while the blue histograms represent the RMSE when the model was fitted using the simulation values.}
	\label{fig:TMBParEstSdObs}
\end{figure}

\begin{figure}[!htbp]
	\includegraphics{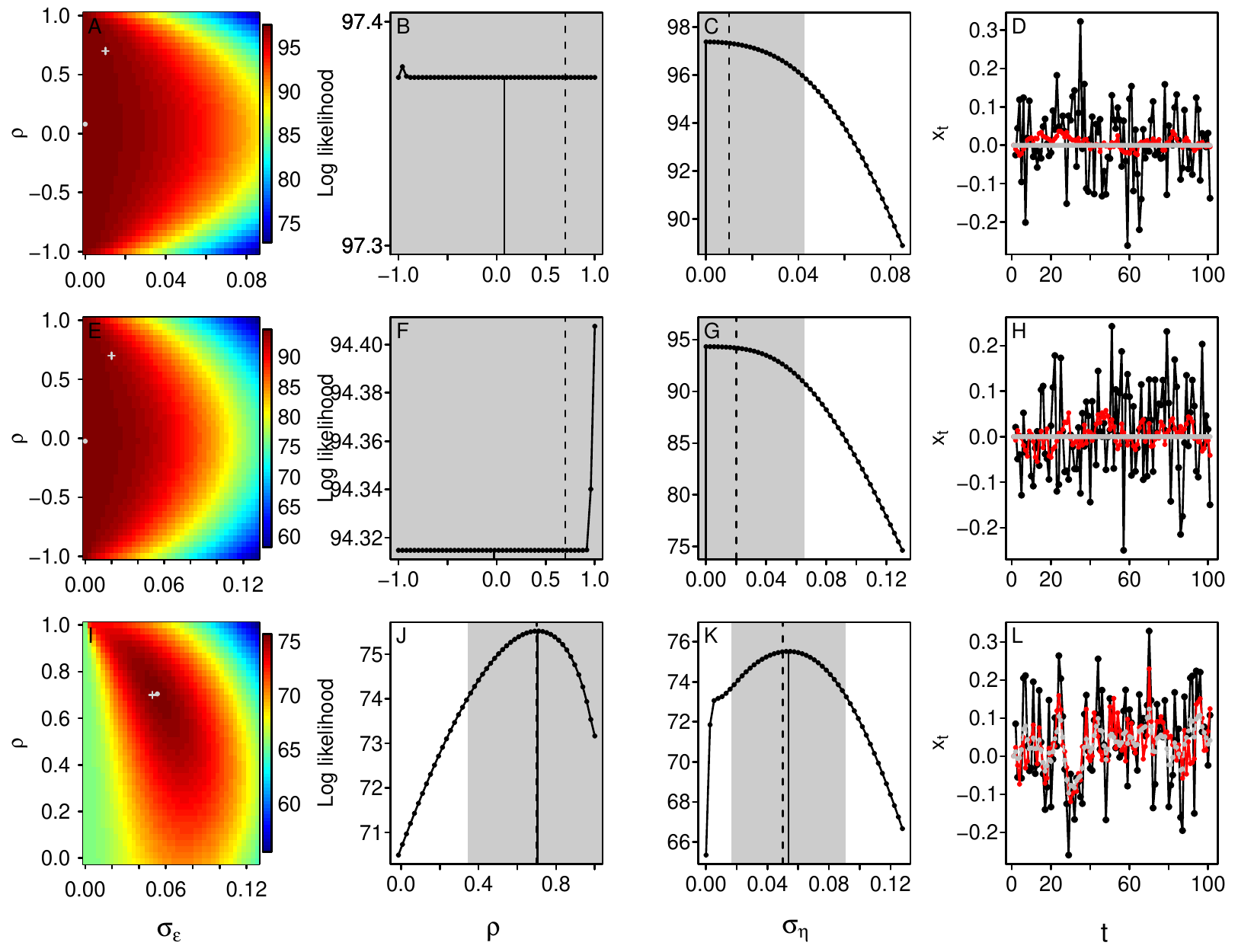}
	\caption{Log likelihood surface and profile for the problematic simulations when the standard deviation of the measurement equation is fixed to the simulated value, $\sigma_{\epsilon} = 0.1$. The first column represents the log likelihood surface for the two estimated parameters, $\mathbf{\theta_{m}} = (\rho, \sigma_{\eta})$. The grey dot is the maximum likelihood estimate and the grey cross is the simulated value. The second and third columns represent the log likelihood profile for $\rho$ and $\sigma_{\eta}$. The curve represents the log likelihood when the focal parameter is fixed (the other parameter are optimise to maximise the log likelihood). The dash lines are the true parameter values (i.e., value used for the simulation), the full lines are the maximum likelihood estimates and the grey bands represent the $95\%$ CI. The last column shows the time-series. The black lines represent the observations, $y_t$, the red lines the simulated true states, $x_t$, and the grey dashed lines the estimated states, $\hat{x}_t$.}
	\label{fig:LProf}
\end{figure}

As mentionned in section \ref{MesRes} of the main text, fixing the standard deviation of the measurement error to the simulated value, $\sigma_{\epsilon} = 0.1$, helped reduce the estimation problems (compare Fig. \ref{fig:TMBParEstSdObs} to Fig. \ref{fig:TMBParEst}). In particular, the state RMSE when the parameters were estimated were much closer to those when the parameters were fixed to the simulated values (e.g., compare Fig. \ref{fig:TMBParEstSdObs}D to Fig. \ref{fig:TMBParEst}D). In this case, only 5.0\% of the simulations had a $\text{RMSE}_{\hat{\theta}}$ value that was $50\%$ larger than their $\text{RMSE}_{\theta}$. In addition, likelihood profiles were more unimodal than when all parameters were estimated (e.g., compare  Fig. \ref{fig:LProf}J to Fig. \ref{fig:LProfAll}J), and the state estimates were no longer simply echoing the observations (e.g., compare Fig. \ref{fig:LProf}L to Fig. \ref{fig:LProfAll}L). However, using measurement error information did not completely resolve the estimation problems. Some parameter estimates continued to be on the boundary of parameter space and far from their simulated values (e.g., Fig. \ref{fig:TMBParEstSdObs}E). In addition, some likelihood profiles remained flat and some CIs spanned the entire parameter space (e.g., Fig. \ref{fig:LProf}B).

\FloatBarrier
\newpage
\setcounter{figure}{0} 
\setcounter{table}{0} 
\section{Diagnostic tools}
\label{App:diagTools}

Identifying whether the model has parameter estimability problems is an important step towards avoiding biased inference. In this Appendix, we present a few diagnostic tools for both the more traditional likelihood-based methods and for Bayesian methods.

One of the best way to check whether a model is capable of estimating the parameters and states is through a simulation study such as the one presented in this manuscript. While an extensive simulation study that investigates a variety of parameter values is necessary to assess the overall capacity of the model, in many cases it may be sufficient to focus on parameter values similar to those estimated from the real data. An example of such focussed simulation study is presented in Appendix \ref{App:PBsim}. One important aspect of these simulation studies is that they require to run a large sample of simulations (we used a sample of 200-500 simulations per parameter set). Using a single simulation can be extremely misleading. For instance, only 29.6\% of our main simulations were problematic (see section \ref{SimRes} from the main text). However, repeatedly fitting a model to multiple time-series may only be feasible with computationally-efficient method. Computing a simulation study with \texttt{rjags} is much less practical than with \texttt{TMB}.

\begin{figure}[!htbp]
	\includegraphics{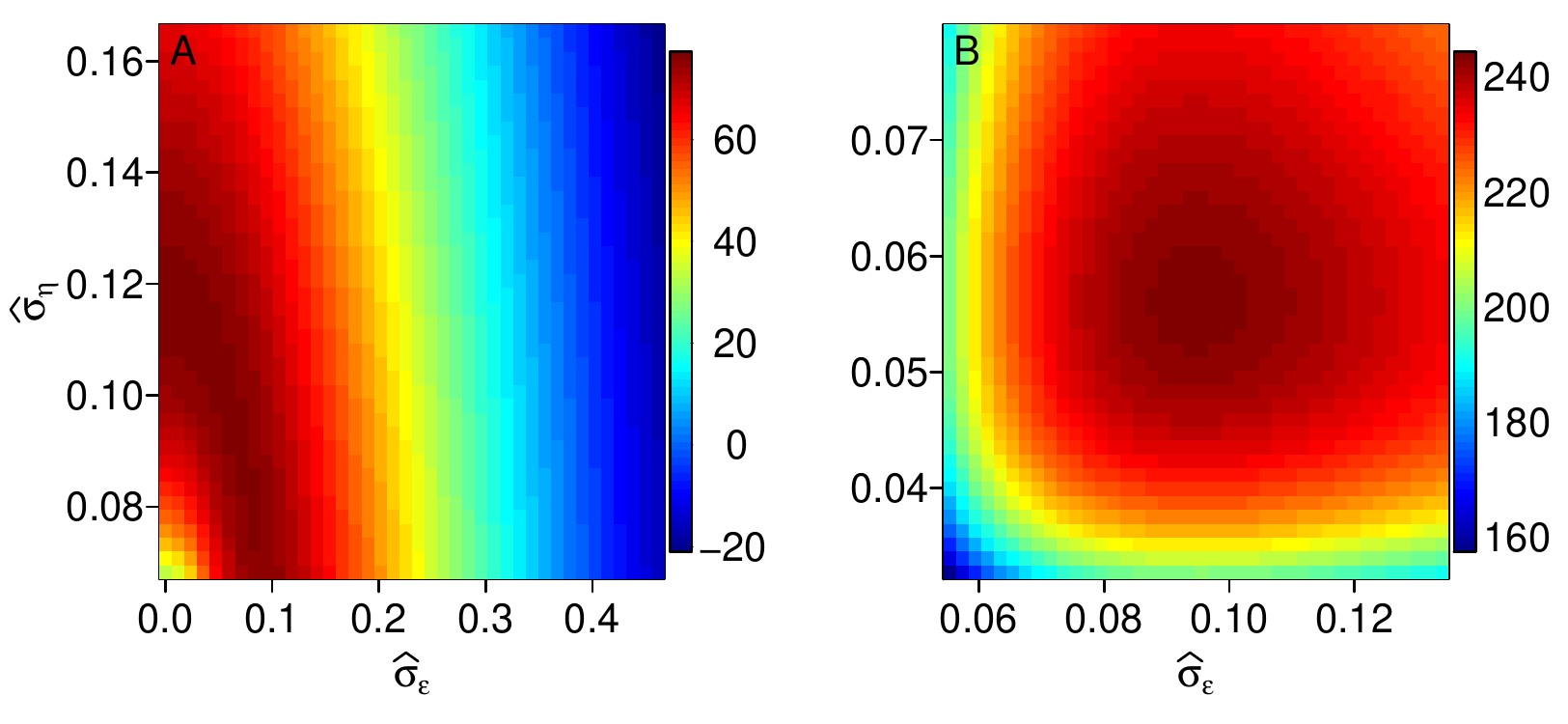}
	\caption{Log likelihood surface for the measurement error and process stochasticity. A) Surface when the states and parameters are estimated. B) Surface when the true states are known.}
	\label{fig:diagTMB}
\end{figure}

As shown repeatedly in the manuscript, one way to identify the potential for estimation problems with likelihood-based methods is by investigating the profile likelihood. Flat, jagged, or bimodal profile likelihoods indicate the potential for parameter-estimation problems (e.g., Fig. \ref{fig:LProfAll}A-C). In contrast, smooth unimodal profile likelihoods, such as those where we know the true states (e.g., Fig. \ref{fig:LProffixx}A-C), indicate that there is no obvious estimation problems. A more comprehensive investigation can be done through the visualization of a likelihood surface \citep{Polansky2009}. For example, by looking at the parameters two-by-two. As an example, we used one of the problematic simulations (i.e., $\text{RMSE}_{\hat{\theta}} > 1.5 \times \text{RMSE}_{\theta}$), where $\sigma_{\eta} = 0.05$. We computed the likelihood surface for the measurement error and process stochasticity. To demonstrate the difference between a problematic and a well-behaved model, we compared the case when the states were estimated (as in the main text) to the case when the states were known (Appendix \ref{App:FixX}). We can see that when the states are estimated the likelihood surface has a diagonal ridge indicating that it is difficult to separate the values of $\sigma_{\eta}$ from those of $\sigma_{\epsilon}$ (Fig. \ref{fig:diagTMB}A). In contrast, in the case when the states are known, we have a well-behaved unimodal likelihood surface (Fig. \ref{fig:diagTMB}B).
 
\begin{figure}[!htbp]
	\includegraphics{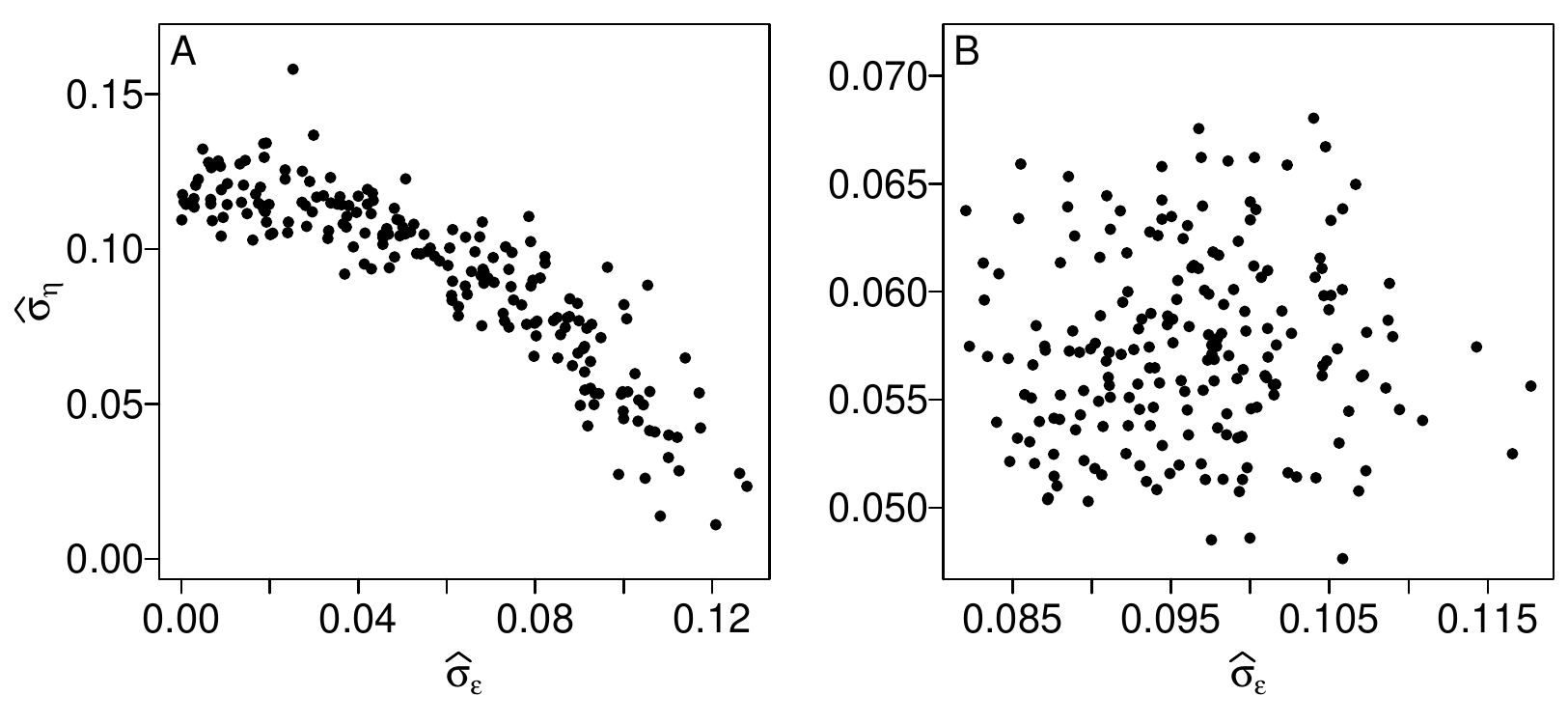}
	\caption{Posterior distribution of the measurement error and process stochasticity. A) Distribution when the states and parameters are estimated. B) Distribution when the true states are known.}
	\label{fig:diagBayes}
\end{figure}

When the model is fitted with Bayesian methods, chain convergence is often used as a diagnostic. The sample paths of MCMC chains for non-identifiable parameters may interchange their values and lead to numerical or convergence problems \citep{Cressie2009}. However, in our case, very few replicates had convergence problems and many of the converged chains lead to biased estimates (see Appendix \ref{App:DLMandJAGS}). To further investigate the potential for estimation problems, in particular, to verify whether the estimates from the different parameters are correlated, one can investigate the posterior distribution of parameters. To show how this method has similarity to investigating the likelihood surface, we used the same example as above. We compared the posterior distribution of model described in eq. \ref{eq:mesEq}-\ref{eq:procEq} (see Appendix \ref{App:DLMandJAGS} for the description of priors) when the states were estimated as opposed to when the states were known. As we can see in Fig. \ref{fig:diagBayes}A, when the states are estimated the posterior distribution of the measurement error and process stochasticity appears strongly correlated, indicating that there is an estimation problem. In contrast, the posterior distribution when the states are known does not appear correlated (Fig. \ref{fig:diagBayes}B), indicating that there is no obvious estimation problem. This is consistent with the results of Appendix \ref{App:FixX}, which shows that when we know the true states, the estimated parameters are close to their true values.

\end{spacing}
\end{document}